\documentclass[11pt,a4paper]{article}

\usepackage[utf8]{inputenc}
\usepackage[T1]{fontenc}
\usepackage{lmodern}
\usepackage{textcomp}
\usepackage[english]{babel}
\usepackage{amsmath, amssymb, mathtools}
\usepackage{braket}
\usepackage{bm}
\usepackage{tikz}
\usepackage{graphicx,caption}
\usepackage[a4paper]{geometry}
\usepackage{footnote}
\usepackage{subcaption}
\usepackage{cite}

\usepackage[unicode=true,
bookmarks=true,bookmarksopen=false,
breaklinks=false,pdfborder={0 0 0},colorlinks=true]
{hyperref}

\usepackage{xcolor}
\definecolor{cblue}{rgb}{0.16, 0.32, 0.75}
\definecolor{cred}{rgb}{0.7, 0.11, 0.11}
\hypersetup{%
	,linkcolor=cred
	,citecolor=cblue
}

\newcommand{\e}{\mathrm{e}}

\newcommand{\dd}{\mathrm{d}}
\newcommand{\ii}{\mathrm{i}}

\renewcommand{\Re}{\operatorname{Re}}
\renewcommand{\Im}{\operatorname{Im}}

\usepackage{etoolbox}
\makeatletter
\def\@mkboth#1#2{}
\newlength\appendixwidth
\preto\appendix{\addtocontents{toc}{\protect\patchl@section}}
\newcommand{\patchl@section}{%
	\settowidth{\appendixwidth}{\textbf{Appendix }}%
	\addtolength{\appendixwidth}{1.5em}%
	\patchcmd{\l@section}{1.5em}{\appendixwidth}{}{\ddt}%
}
\makeatother

\usepackage{perpage}
\MakePerPage{footnote}

\begin{document}

	\begin{center}
	\LARGE
	\textbf{Stationary excitation waves and multimerization \\ in arrays of quantum emitters} \par \bigskip
	
	\normalsize
	Davide Lonigro\textsuperscript{1,2,*},  
	Paolo Facchi\textsuperscript{1,2}, 
	Saverio Pascazio\textsuperscript{1,2},\\
    Francesco V. Pepe\textsuperscript{1,2}, 
	Domenico Pomarico\textsuperscript{3} \par \bigskip
	
	\textsuperscript{1}\footnotesize Dipartimento di Fisica and MECENAS, Universit\`a di Bari, I-70126 Bari, Italy \par
	\textsuperscript{2}\footnotesize INFN, Sezione di Bari, I-70126 Bari, Italy \par
	\textsuperscript{3}\footnotesize Struttura Semplice Dipartimentale di Fisica Sanitaria, \\I.R.C.C.S. Istituto Tumori ``Giovanni Paolo II'', I-70124 Bari, Italy\par
	\textsuperscript{*}\footnotesize Corresponding author: \texttt{davide.lonigro@ba.infn.it}
	\par \bigskip
	
	\date{}
\end{center}

\begin{abstract}
	We explore the features of an equally-spaced array of two-level quantum emitters, that can be either natural atoms (or molecules) or artificial atoms, coupled to a field with a single continuous degree of freedom (such as an electromagnetic mode propagating in a waveguide). We investigate the existence and characteristics of bound states, in which a single excitation is shared among the emitters and the field. We focus on bound states in the continuum, occurring in correspondence of excitation energies in which a single excited emitter would decay.
	We characterize such bound states for an arbitrary number of emitters, and obtain two main results, both ascribable to the presence of evanescent fields. First, the excitation profile of the emitter states is a sinusoidal wave. Second, we discuss the emergence of multimers, consisting in subsets of emitters separated by two lattice spacings in which the electromagnetic field is approximately vanishing.
\end{abstract}

\section{Introduction}

Cooperative effects play an important role in the behavior and time evolution of quantum systems. The emission of light by well-separated atoms and molecules is uncorrelated, so that the states of a single quantum emitter determines the main characteristics of radiation. On the other hand, when quantum interference effects become important, typically in closely packed ensembles of emitters, cooperative effects take over and drastically modify the features of the emitted radiation. 
Typically, such effects compete with the dephasing induced by the dipole-dipole interactions. 

In this scenario, among the most interesting phenomena there are certainly super- and sub-radiance, by which the spontaneous emission of radiation in a transition between two atomic levels leads to coherent emission by an atomic ensemble, enhancing~\cite{dicke,SRreview,benedict} or reducing \cite{Fedorov1} the 
decay rate. 
Clearly, these effects bear a profound influence on the formulation of the state of the radiating system. 
Since super- and sub-radiance are general phenomena, that can be observed also in large quantum systems, such as artificial atoms, quantum dots~\cite{SSWFBPH} and superconducting circuits \cite{hoi}, we will keep our discussion as general as possible, and refer generically to quantum ``emitters'' of radiation.

On the other hand, light scattering by cold atomic clouds can induce photon-mediated effective long-range interactions between atoms, leading to cooperative effects even at low atomic densities. While superradiance has been extensively studied since Dicke's seminal proposal~\cite{dicke},
subradiance in large cold atom clouds has been observed only rather recently \cite{guerin16,weiss18}. Although these phenomena take place at light wavelengths that are typically much larger than interatomic distances~\cite{Kaiser1,Kaiser2}, a number of interesting quantum resonance effects appear at shorter wavelengths, comparable to the distance among atoms.

The present investigation pertains to the latter regime. Recent advances in quantum technologies have made possible the realization of novel experimental platforms, in which light propagation is (effectively) one-dimensional. These platforms make use of a wide variety of coherent quantum systems, ranging from optical fibers~\cite{onedim3,onedim4}, cold atomic systems~\cite{focused1,focused2,focused3}, and superconducting qubits~\cite{onedim5,onedim6,mirror1,mirror2,atomrefl1,leo5}, to photonic crystals~\cite{kimble1,kimble2,onedim1,onedim2,ck}, and quantum dots in photonic nanowires~\cite{semiinfinite1,semiinfinite2}.

One of the most interesting features of such 1D systems is the presence of single-excitation bound states in the continuum (BICs) \cite{tanaka,longhi2007}. Such states represent an extreme case of subradiance, where the excitation is shared in a stable way between the emitters and the field, so that radiation is completely trapped by the emitters even though the energy would be sufficient to yield photon propagation. The physical characteristics of quantum emitters in waveguides has been studied in a variety of situations, for single~\cite{focused1,mirror2,boundstates1,threelevel} as well as double and multiple emitters
\cite{refereeA1,refereeA2,PRA2016,oscillators,baranger,baranger2013,NJP,yudson2014,laakso,pichler,Fedorov1,PRA2018,pichler2,bello,bernien,dong,fang,fang14,goban,ck,gu,guimond,lalumiere,lodahl,paulisch,ramos14,ramos,cirac1,tsoi,yudsonPLA,yudson2008,calajo15}. In all cases, the quantum correlations between the emitters are due to the 1D photonic field.

The case of $n=3$ and $n=4$ emitters, analyzed via a non-perturbative treatment based on the Friedrichs-Lee model~\cite{lee,fridleeold,fridleeold2,singcoup}, was studied in~\cite{bic}. The role of non-perturbative photon-mediated interactions between emitters was shown to be crucial to find the correct bound states.

In this article, we will apply this formalism to find an explicit expression for the BICs for an arbitrary number $n$ of emitters. We will focus on the global state of the emitters and will show that their excitation amplitude profile follows a sinusoidal law: single-excitation BICs are therefore \textit{excitation waves}, conceptually analogous to the stationary spin waves emerging in 1D magnetic models \cite{bloch,holstein,freeman}. We will also unearth the presence of \emph{multimerization}: under suitable conditions, the excitation amplitude displays a modular repetition of quasi-decoupled components. One might expect that multimers arise independently of each other: by contrast, we shall see that the evanescent fields that are present between adjacent multimers play a crucial role. Our procedure will be based on a nearest-neighbor approximation for the inter-emitter interactions and will hold under very general assumptions.

The paper is structured as follows. In Section~\ref{sec:model} we present the model, introduce its propagator matrix, and discuss its role in the evaluation of the BICs. In Section~\ref{sec:evaluation} we determine the structure of the BICs for an arbitrary number of emitters. Finally, in Section~\ref{sec:multimerization}, we discuss the emergence of multimerized states and find a condition for constructing new BICs by arranging together smaller building blocks.

\section{The model} \label{sec:model}

In this section, after first introducing in Subsection~\ref{subsec:generalities} the basic features of the model to be investigated, we show in Subsection~\ref{subsec:eigenproblem} how the BICs in the single-excitation sector can be conveniently described via the propagator matrix. The method outlined in this section will be then applied in Section~\ref{sec:evaluation}.

\subsection{Generalities}\label{subsec:generalities}
\begin{figure}\centering
	\includegraphics[scale=0.45]{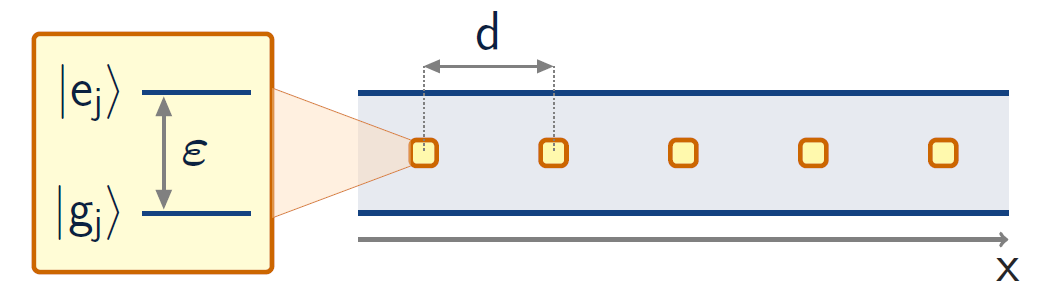}
	\captionsetup{justification=justified, singlelinecheck=false}
	\caption{Schematic representation of the physical system: an array of two-level atoms, labelled by the index $j$, are characterized by a ground state $\ket{g_j}$ and an excited state $\ket{e_j}$, separated by the energy $\varepsilon$. The atoms are placed at equal distances $d$ and coupled to a linear waveguide mode.}
	\label{fig:systemn}
\end{figure}

We consider a system of $n$ identical two-level emitters, with excitation energy $\varepsilon$, equally spaced at a distance $d$ (see Fig.~\ref{fig:systemn}), with ground and excited states $\ket{g_j}$ and $\ket{e_j}$, coupled to a bosonic field with energy profile $\omega(k)$. 
The Hamiltonian reads
\begin{equation}
\label{tothamiltonian}
H= H_0+ H_{\mathrm{int}}
\end{equation} 
where 
\begin{equation}
\label{freehamiltonian}
H_0  =   \varepsilon \sum_{j=1}^n \sigma^+_j \sigma^-_j  + \int \dd k\;\omega(k)b^{\dagger}(k) b(k) 
\end{equation} 
is the free Hamiltonian and 
\begin{equation}\label{Hint}
H_{\mathrm{int}} = \sum_{j=1}^n \int \dd k \Bigl[ F(k)\,\e^{-\ii (j-1)kd} \sigma^-_j \otimes 
b^\dag(k)  +  \mathrm{H.c.} \Bigr] 
\end{equation}
the interaction. In the above formulas, $\sigma^-_j=\ket{g_j}\!\bra{e_j}$ and $\sigma^+_j=\ket{e_j}\!\bra{g_j}$ are the lowering and raising operators of the $j$-th emitter, and $b^\dag(k)$, $b(k)$ are the photon creation and annihilation operators, satisfying the canonical commutation relations $[b(k),b(k')]=0$, $[b(k),b^{\dagger}(k')]=\delta_{kk'}$.
The interaction has a rotating-wave form, with $F(k)$ being the form factor that describes the coupling of the emitters with a boson of momentum $k$.\footnote{In general, the coupling between each emitter and the field may be described by a coupling function $F_j(k)$, but, since the emitters are identical, $F_j(k)=F(k)\,\e^{-\ii (j-1)kd}$. Indeed, in the position representation the coupling function of the $j$th emitter, placed at  position $x_j$ in the waveguide, must be given by $\mathcal{F}_j(x)=\mathcal{F}(x-x_j)$ for some function $\mathcal{F}(x)$. In our case, $x_j=(j-1)d$ and a Fourier transform yields the result.}

In order to proceed, some assumptions about the functions $\omega(k)$ and $F(k)$ will be needed. A concrete example is represented by an array of two-level atoms coupled with a single transverse mode of the electromagnetic field in a linear waveguide. We assume a dispersion relation of the form 
\begin{equation}
\omega(k)=\sqrt{k^2+m^2}, 
\label{eq:KleinGordon}
\end{equation}
expressed in natural units. In the case of a waveguide with rectangular cross section, whose sides are related by $L_z<L_y$, the ``photon mass'' $m$ is proportional to the inverse smaller size $L_z^{-1}$ \cite{jackson}. The form factor in this case reads
\begin{equation}\label{eq:formfactor}
F(k) = \sqrt{\frac{\gamma}{2\pi\omega(k)}},
\end{equation}
where $\gamma>0$ is a coupling constant. We shall focus henceforth on this case. However, as shown in Appendix~\ref{app:self}, the results that we shall discuss in the following turn out to be largely independent of the particular form of the coupling $F(k)$ and the dispersion relation $\omega(k)$. We remark that, in order to simplify the computation, it is a common choice to replace $F(k)$ with a constant form factor, and assign linear dispersion relations to left- and right-propagating photons (see, e.g., Ref.~\cite{shen2009,yudson2014}). Although such an approximation
is able to describe bound states in the continuum, it fails to capture non-Markovian effects \cite{dinc2019}, such as those which constitute the most relevant phenomenology presented in this work. 

Observe that the rotating-wave form of the interaction Hamiltonian entails that the total number of excitations 
\begin{equation}
\mathcal{N}= \sum_{j=1}^n \sigma^+_j \sigma^-_j + \int \dd k \; b^{\dagger}(k) b(k)
\end{equation}
is conserved, $[H,\mathcal{N}]=0$. 
In the single-excitation sector $\mathcal{N}=1$, the state vectors read
\begin{equation}\label{state1}
\ket{\Psi} = \sum_{j=1}^n a_j \ket{e_j} \otimes \ket{\mathrm{vac}} + \ket{g} \otimes \int \dd k\, \tilde{\xi}(k)\, b^{\dagger}(k) \ket{\mathrm{vac}} ,
\end{equation}
where $\ket{\mathrm{vac}}$ is the field vacuum state,
\begin{equation}
\ket{g} =   \ket{g_1} \otimes  \ket{g_2} \otimes \cdots \otimes  \ket{g_n}, \qquad  \ket{e_j} = \sigma_j^+ \ket{g},
\qquad \tilde{\xi}(k) = \int \frac{\mathrm{d}x}{\sqrt{2\pi}}\: \xi(x)\, \e^{-\ii k x},
\end{equation}
and $\bm{a} \in\mathbb{C}^n$, $\xi \in L^2(\mathbb{R})$ are constrained by the normalization condition
\begin{equation}\label{eq:normalized}
	\sum_{j=1}^n|a_j|^2+\int|\xi(x)|^2\,\mathrm{d}x=1.
\end{equation}
Here, the vector $\bm{a}=(a_1,\dots,a_n)^\intercal$ is the excitation amplitude profile of the emitters and $\xi(x)$ the photon wavefunction.

\subsection{Propagator and bound states}\label{subsec:eigenproblem}
The single-excitation sector was investigated for $n=2,3,4$ atoms in Refs.~\cite{PRA2016,oscillators} and shown to contain, for generic dispersion relations and form factors 
(in particular, for those given in Eqs.~\eqref{eq:KleinGordon}--\eqref{eq:formfactor}) and for selected interatomic distances, nontrivial atom-photon bound states. Our main objective is to extend these results to arbitrary $n$ and unearth genuine collective effects.

Bound states are given by the solutions of the equation
\begin{equation}\label{eq:eigenvalueeq}
H\ket{\Psi}=E\ket{\Psi}
\end{equation} 
for some real energy $E$, with $\ket{\Psi}$ as in Eq.~\eqref{state1} and normalized as in Eq.~\eqref{eq:normalized}, for some $\bm{a}$ and $\xi(x)$. In particular, bound states with energy $E>\omega(0)$ are bound states in the continuum (BICs), since they are embedded in the continuous set of the field frequencies. 

As shown in Appendix~\ref{sec:eigenvalue}, the content of Eq.~\eqref{eq:eigenvalueeq} can be rephrased as follows. We consider the \textit{propagator} matrix of the model, which we compute in Appendix~\ref{app:self}:
\begin{align}\label{eq:propagator}
\mathrm{G}_{j\ell}^{-1}(E)&=(\varepsilon-E)\delta_{j\ell} +\int \dd k\;\frac{|F(k)|^2}{E+\ii0-\omega(k)}\e^{\ii (j-\ell)kd}\nonumber\\
&=(\varepsilon-E)\delta_{j\ell}- 2\pi\ii \frac{|F(k(E))|^2}{\omega'(k(E))}\e^{\ii |j-\ell|k(E)d}+\beta_{j-\ell}(E),
\end{align}
where $k(E)$ is the positive solution of $\omega(k)=E$, i.e.\ the momentum of a photon with energy $E$, and $\beta_{j}(E)$ a real-valued function satisfying
\begin{equation}\label{eq:small1}
\left|\beta_j(E)\right|\leq e^{- |j| m d}\left|\beta_0(E)\right|.
\end{equation}

Interestingly, all terms $\beta_{j}(E)$ would vanish if we would chose a linear 
dispersion relation 
and a constant form factor, 
namely $\omega(k) =\omega_0+v(k-k_0)$ and $F(k) = F_0$. However, any more accurate physical model with a correction to the linear-dispersion-relation approximation and/or the flat-coupling approximation will give nonvanishing values of $\beta_j(E)$.  
This is a crucial point: despite these terms being (for $j\neq0$) exponentially small in $md$, they will play a fundamental role in determining the very structure of the BICs, as we will see in the next sections. 

Then, a real energy $E$ is an eigenvalue in Eq.~\eqref{eq:eigenvalueeq} if
\begin{equation}\label{eq:BICs1}
\det\mathrm{G}^{-1}(E)=0;
\end{equation}
this energy will corresponds to a BICs if $E>\omega(0)$. As shown in Appendix~\ref{sec:eigenvalue} and~\ref{sec:phwave}, a corresponding eigenstate $\ket{\Psi}$ is such that
\begin{itemize}
\item the vector of atomic excitation amplitudes $\bm{a} =(a_1,a_2,\dots,a_n)^\intercal$ is a solution of the matrix equation
\begin{equation}\label{eq:BICs2}
\mathrm{G}^{-1}(E)\,\bm{a} = 0 ,
\end{equation}
with the additional constraint
\begin{equation}
F(k(E))\sum_{j=1}^na_j\,\e^{\pm\ii (j-1)k(E) d}=0;
\label{eq-physicalcondition}
\end{equation}
\item the wavefunction $\xi(x)$ is
\begin{align}
\xi(x)&=\frac{1}{\sqrt{2\pi}}\int\dd k\:\frac{F(k)}{E-\omega(k)}\sum_{j=1}^{n}a_j\:\e^{\ii (x-(j-1)d)k}\nonumber\\
&=\sum_{j=1}^n a_j\:\xi_1\bigl(x-(j-1)d\bigr),
\label{eq:wav}
\end{align}
\end{itemize}
where
\begin{equation}\label{eq:singleboson}
\xi_1(x)= \sqrt{2\pi} \frac{F(k(E))}{\omega'(k(E))}  \sin \bigl( k(E) | x |\bigr) + \eta(x),
\end{equation}
with the correction $\eta(x)$ being real and satisfying 
\begin{equation}\label{eq:eta}
|\eta(x)|\leq \e^{-m |x|}|\eta(0)|.
\end{equation}

Eq.~\eqref{eq:wav} shows that the photon wavefunction $\xi(x)$ corresponding to a BIC with energy $E$ is a linear combination of $n$ copies of $\xi_1(x)$, each centered at one of the emitters, $x_j=(j-1)d$: the contribution of each emitter is weighted by the corresponding component of $\bm{a}$. Again, the term $\eta(x)$, which physically corresponds to a small evanescent field, would again vanish if we would take a linear approximation of the dispersion relation. All relevant quantities are summarized in Table~\ref{tab:summary}.

\begin{table}[!htb]\centering
	\begin{tabular}{|c|c|c|}
	\hline 
	\textbf{Quantity}	& \textbf{Description} &  \textbf{Definition}\\ 
	\hline 
	$\varepsilon$	& Excitation energy of the emitters &  --\\
	\hline 
	$d$	& Distance between consecutive emitters &  --\\
	\hline 
	$\omega(k)$	& Dispersion relation of the boson continuum &  \eqref{eq:KleinGordon}\\ 
	\hline 
	$m$	& Lower bound of the continuum (mass cutoff)&  $m=\omega(0)$\\ 
	\hline 
	$F(k)$	& Form factor &  \eqref{eq:formfactor}\\	
	\hline 
	$k(E)$	& Positive solution of $\omega(k)=E$ &  --\\	
	\hline 
$\mathrm{G}^{-1}(E)$, $\mathrm{G}_{j\ell}^{-1}(E)$	& Propagator of the model, and its $(j,\ell)$th element &  \eqref{eq:propagator}\\ 
	\hline 
	$\beta_{j-\ell}(E)$& Correction to $\mathrm{G}^{-1}_{j\ell}(E)$ &  \eqref{eq:B25}\\
	\hline 
$\xi_1(x)$	& Single-emitter wavefunction & \eqref{eq:singleboson} \\ 
	\hline 
$\eta(x)$	& Correction to $\xi_1(x)$ & \eqref{eq:defeta} \\ 
	\hline 
\end{tabular}
\caption{Summary table of the quantities introduced in Section~\ref{sec:model}.}\label{tab:summary}
\end{table}

Summing up, in order to compute the BICs emerging in our model, we need to solve Eqs.~\eqref{eq:BICs1} and~\eqref{eq:BICs2}, with the propagator matrix being given by Eq.~\eqref{eq:propagator}; once we solve this problem, thus finding the admissible energies $E$ of the BICs and the corresponding excitation profiles $\bm{a}$, the associated photon wavefunction is given by Eq.~\eqref{eq:wav}. The next section will be devoted to this problem.

\section{Structure of the bound states in the continuum} \label{sec:evaluation}

The evaluation of BICs for an array of quantum emitters was performed, under less general hypotheses, in Ref.~\cite{PRA2016} for $n=2$ emitters and in Ref.~\cite{bic} for $n=3,4$ emitters. When $n>4$, the eigensystem becomes more and more involved because of the presence of the terms $\beta_{j-\ell}(E)$ in the expression of the propagator, whose number grows with $n$.

At the physical level, these terms may interpreted as (field-mediated) inter-emitter interaction terms. Even though these terms are exponentially suppressed in $md$, and hence ``small'' in the physically interesting regime $md\gg1$, their role turns out to be fundamental: without these contributions the eigensystem would be \textit{degenerate} and no preferred set of stable emitter configurations can be found, while, when taking into account the additional terms, such degeneracy is lifted.

\begin{figure}[!htb]\centering
\includegraphics[scale=0.35]{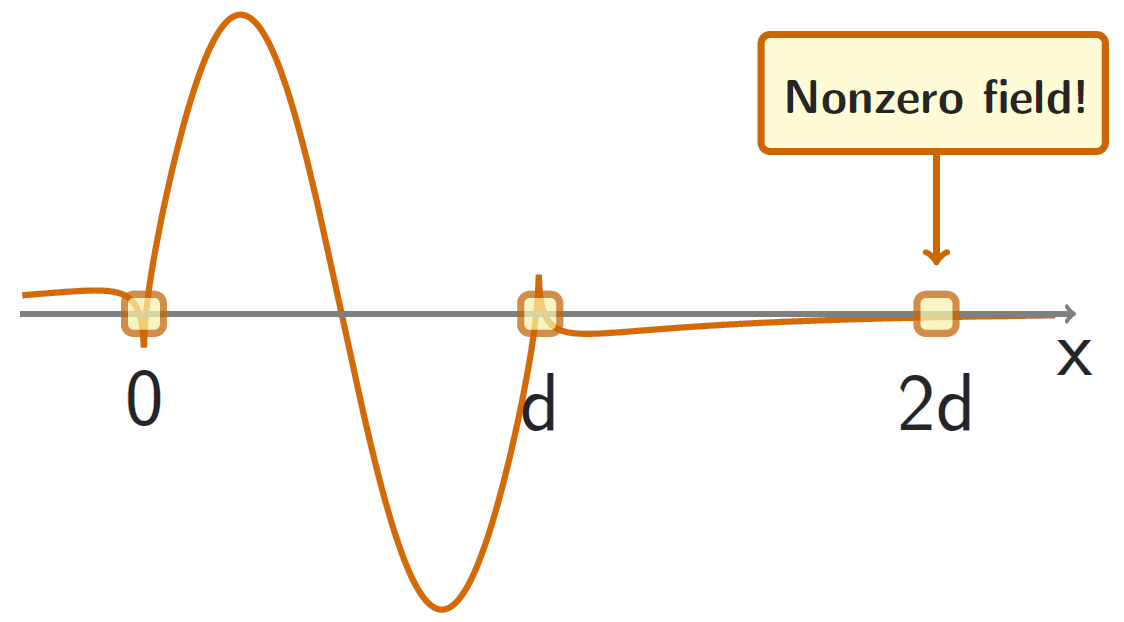}
\captionsetup{justification=justified, singlelinecheck=false}
\caption{An example: configuration with $n=3$ emitters in positions $x=0,d,2d$, in which $a_1=a_2$, $a_3=0$ and the oscillating part of the field is confined between the atoms that share the excitation. 
If there were no evanescent contributions, the field would be exactly confined between the first two emitters and the resulting two-emitter state would be stable, regardless of the presence of the third emitter at $x=2d$. However, the presence of an evanescent field, which couples with the third emitter, causes instability, as it tends to transfer the excitation to the third emitter, and eventually to decay.}
\label{fig:tails}
\end{figure}

This phenomenon can be better understood when looking at the photon wavefunction associated with a given emitter configuration: as shown in Eq.~\eqref{eq:singleboson}, the latter will be the sum of a dominant sinusoidal contribution plus an evanescent contribution $\eta(x)$, given by Eq.~\eqref{eq:eta}. Albeit small, such tails are not confined and induce field-mediated coupling among all emitters, which causes instability unless particular configurations are chosen. An example is displayed in Fig.~\ref{fig:tails}: emitter configurations in which only two neighboring emitters resonate, with the other ones ``switched off'', are \textit{not} allowed even though the state would be stable in the absence of the other emitters.

Therefore, a generic BIC in a multi-atom array is expected to be a \textit{collective} state: the excitation on the BICs must be distributed in a very specific way along the whole array, and must be chosen in such a way that the effects of all such fields even out, yielding a stable state. 

In the absence of evanescent field one would have a manifold of degenerate \textit{local} solutions with the same energy $E$, consisting of radiation trapped between pairs of emitters, that could be pasted together in an arbitrary way. However, in the presence of evanescent fields, most of these states are unstable, and only a few selected ones emerge as stable configurations. The role of the evanescent fields is to \emph{select} collective stable states. In particular, even those emitters (if any) which have zero amplitude and thus are decoupled from the field must be placed at specific positions along the chain (namely, the nodes of the field), and, in this sense, they also cooperate to yield the overall stability of the state. We will see this selection mechanism at work in stabilizing multimer configurations in Sec.~\ref{sec:multimerization}.

The heuristic discussion above will be supported by calculations. Collective configurations will emerge as a natural consequence of the structure of the propagator: we will prove that the atomic excitation amplitudes in a stable configuration are interpolated by (either exact or ``deformed'') sinusoidal functions, providing an interesting analogy with the stationary spin waves that emerge in the Heisenberg model~\cite{bloch,holstein,freeman}.

The present section is organized as follows:
\begin{itemize}
	\item in Subsection~\ref{subsec:nearest} we present the idea on which the computation of BICs is based;
	\item in Subsection~\ref{subsec:deform} we will sum up the evaluation of the energies of all possible BICs and their corresponding excitation profiles $\bm{a}$, obtained as solutions of Eq.~\eqref{eq:BICs2}, with all details being reported in~Appendix~\ref{app:calculation};
	\item in Subsection~\ref{subsec:results}, after computing the photon wavefunctions $\xi(x)$ corresponding to such profiles via Eq.~\eqref{eq:wav}, we discuss our findings.
\end{itemize}
A particular feature of some BICs that comes out as a direct consequence of our discussion, namely the presence of multimerized BICs, will be investigated in Section~\ref{sec:multimerization}.

\subsection{A nearest-neighbor approximation}\label{subsec:nearest}
We shall search for bound states in the continuum with values of energy (close to) $E=E_\nu$, where we define the $\nu$th \textit{resonant energy} by
\begin{equation}\label{eq:resonance0}
	E_\nu=\omega\left(\frac{\nu\pi}{d}\right),\quad\nu\in\mathbb{N}.
\end{equation}
This choice is motivated by the following observation: if we were to discard all terms $\beta_{j-\ell}(E)$ with $j\neq\ell$ in the propagator \eqref{eq:propagator}, Eq.~\eqref{eq:BICs1} would be satisfied if and only if $E=E_\nu$ for some $\nu$, and thus all BICs would be found at such values of energy. Correspondingly, the system would exhibit an $(n-1)$-dimensional degenerate space of BICs with energy $E=E_\nu$. This phenomenon is extensively discussed in Ref.~\cite{bic} and, for completeness, revised in Appendix~\ref{subsec:degen}. As we have seen above, this is the case for a linear dispersion relation and a flat form factor, when all $\beta_j(E)$ for $j\neq0$ vanish. 

The degenerate situation outlined above is drastically modified when one takes into account the full structure of the propagator, which is the case when one goes beyond the linear approximation. This problem was already analyzed for the cases of $n=3$ and $n=4$ emitters \cite{bic}, where it was shown that the presence of nonvanishing off-diagonal terms $\beta_{j-\ell}$, no matter how small, lifts the degeneracy: the available BICs, with energy either equal or close to $E_\nu$, emerge for distinct, albeit close, values of $\varepsilon$. This means that the system exhibits at most \textit{one} truly stable BIC. Besides, as discussed in Appendix~\ref{subsec:parity}, BICs turn out to have a well-defined parity.

When $n>4$, the analytic study of the full structure of the propagator becomes unfeasible, and therefore we need a way to extract information about the resonant degeneracy breaking patterns, while keeping calculations viable. The basic idea is the following: since
\begin{equation}
\frac{\beta_j(E)}{\beta_0(E)} = O\bigl(\e^{-|j|md}\bigr),
\end{equation}
the off-diagonal terms with $|j-\ell|>1$ are of higher order in $\e^{-md}$ than those with $|j-\ell|=1$, and therefore, in order to study the degeneracy lifting at large $md$, it is enough to consider the first-order approximation
\begin{equation} \label{NNint}
\beta =(\beta_1, \dots,\beta_{n-1}) \sim (\beta_1, 0, \dots, 0).
\end{equation}
Indeed, we will now show that the presence of the largest of such contributions, namely $\beta_1$, is sufficient to fully remove the $n-1$ degeneracy and thus determine the nondegenerate BICs in the large $md$ regime. Physically, since $\beta_{j-\ell}$ is related to a field-mediated interaction involving the $j$th and $\ell$th emitter, the replacement \eqref{NNint} may be regarded as a \textit{nearest-neighbor approximation}. Higher-order corrections in $md$ (that is, terms $\beta_{j-\ell}(E)$ with $|j-\ell|>1$) would only contribute by $O(\e^{-2md})$ corrections, without changing the qualitative picture that we will outline.

\subsection{Exact and deformed excitation waves}\label{subsec:deform}
In the approximation discussed above, the propagator \eqref{eq:propagator} of the model, evaluated at a resonant energy $E_\nu$~\eqref{eq:resonance0}, can be written in a convenient form. By directly substituting $E=E_\nu$ in \eqref{eq:propagator}, we get
\begin{equation}\label{eq:propagator2}
	\mathrm{G}^{-1}(E_\nu)=\mathrm{G}^{-1}_0(E_\nu)-\ii\,\beta_1(E_\nu)\Delta_n,
\end{equation}
where
\begin{itemize}
	\item $\mathrm{G}^{-1}_0(E_\nu)$ is the zeroth order approximation of the propagator, that is, the one that would be obtained by discarding \textit{all} terms $\beta_{j-\ell}$ with $j\neq\ell$. By Eq.~\eqref{eq:propagator}, it reads
	\begin{equation}\label{eq:propagator0}
\mathrm{G}^{-1}_0(E_\nu)=\left(\varepsilon-E_\nu+\beta_0(E_\nu)\right)\mathrm{I}_n-2\pi\ii\frac{|F(k(E_\nu))|^2}{\omega'(k(E_\nu))}\bm{u}_\nu\bm{u}_\nu^\intercal,
	\end{equation}
	with $\bm{u}_\nu =\bigl(1,(-1)^\nu,1,(-1)^\nu,\dots\bigr)^\intercal$.
	\item $\Delta_n$ is the $n\times n$ matrix defined by
	\begin{equation}\label{eq:laplace}
	\Delta_n=\begin{pmatrix}
	0&1&&&\\
	1&0&1&&\\
	&1&0&1&&&\\
	&&\ddots&\ddots&\ddots\\
	&&&1&0&1&\\
	&&&&1&0&1\\
	&&&&&1&0
	\end{pmatrix},
	\end{equation}
	and corresponds to the adjacency matrix (discrete Laplacian) of a one-dimensional regular graph~\cite{graph, ChuYau} with Dirichlet boundary conditions. Its main properties are recalled in~Appendix~\ref{app:laplace}.
\end{itemize}
Consequently, when taking into account the role of the terms $\beta_1(E)$ in the propagator (physically, nearest-neighbor photon-mediated interactions between the emitters), we are perturbing the original propagator $\mathrm{G}_0^{-1}(E)$ with a multiple of the matrix $\Delta_n$. Despite the latter term being exponentially small in $md$ (recall Eq. \eqref{eq:small1}), it does crucially affect the solutions of Eq.~\eqref{eq:BICs1}, since it lifts the degeneracy of the original problem and allows us to individuate an unambiguous set of solutions each corresponding to a BIC emerging at an energy (close to) $E_\nu$ when the excitation energy $\varepsilon$ has a certain value. 

This problem is studied in Appendix~\ref{subsec:nondegen}, to which we refer for details. We will present here the results. First of all, for all $j=1,\dots,n$, define the quantities $\chi^{(j)}$ and $\bm{a}^{(j)}$ via
\begin{equation}\label{eq:eigvl}
\chi^{(j)}=
2\cos\left(\frac{\pi j}{n+1}\right),\qquad
a^{(j)}_\ell=\frac{1}{\mathcal{N}_j}\sin\left(\frac{j\ell\pi}{n+1}\right),
\end{equation}
where $\mathcal{N}_j$ is a normalization constant. These are the eigenvalues and eigenvectors of the matrix $\Delta_n$, as shown in~Appendix~\ref{app:laplace}; in particular, the eigenstates of $\Delta_n$ can be obtained by a sampling at the equally-spaced points $x_\ell = (\ell-1) d$ of sine functions with wavelength 
\begin{equation}\label{eq:lambda}
\lambda_j = \frac{2(n+1)d}{j} .
\end{equation}
We will refer to the eigenvectors $\bm{a}^{(j)}$ of $\Delta_n$ as \textit{excitation waves}. In analogy with the two branches of phonon modes in lattices, we will label excitation waves with low and high frequency as \emph{acoustic} (consecutive amplitudes in phase)  and \emph{optical} modes (consecutive  amplitudes in phase opposition), respectively.

\begin{figure*}[h]
	\centering
	\begin{subfigure}[t]{0.5\linewidth}
		\includegraphics[scale=0.4]{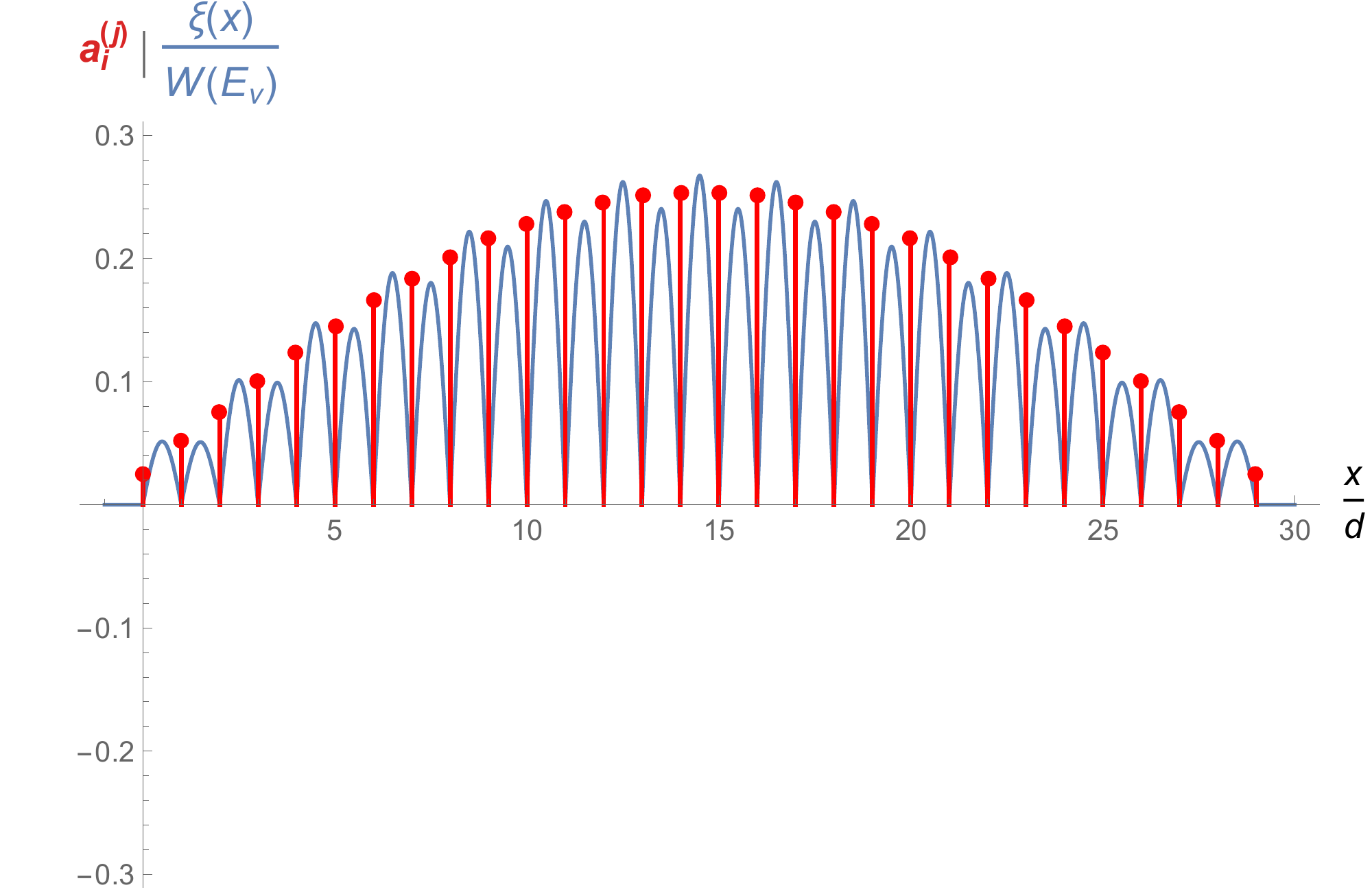}
		\caption{$j=1$}
	\end{subfigure}%
	\begin{subfigure}[t]{0.5\linewidth}
		\includegraphics[scale=0.4]{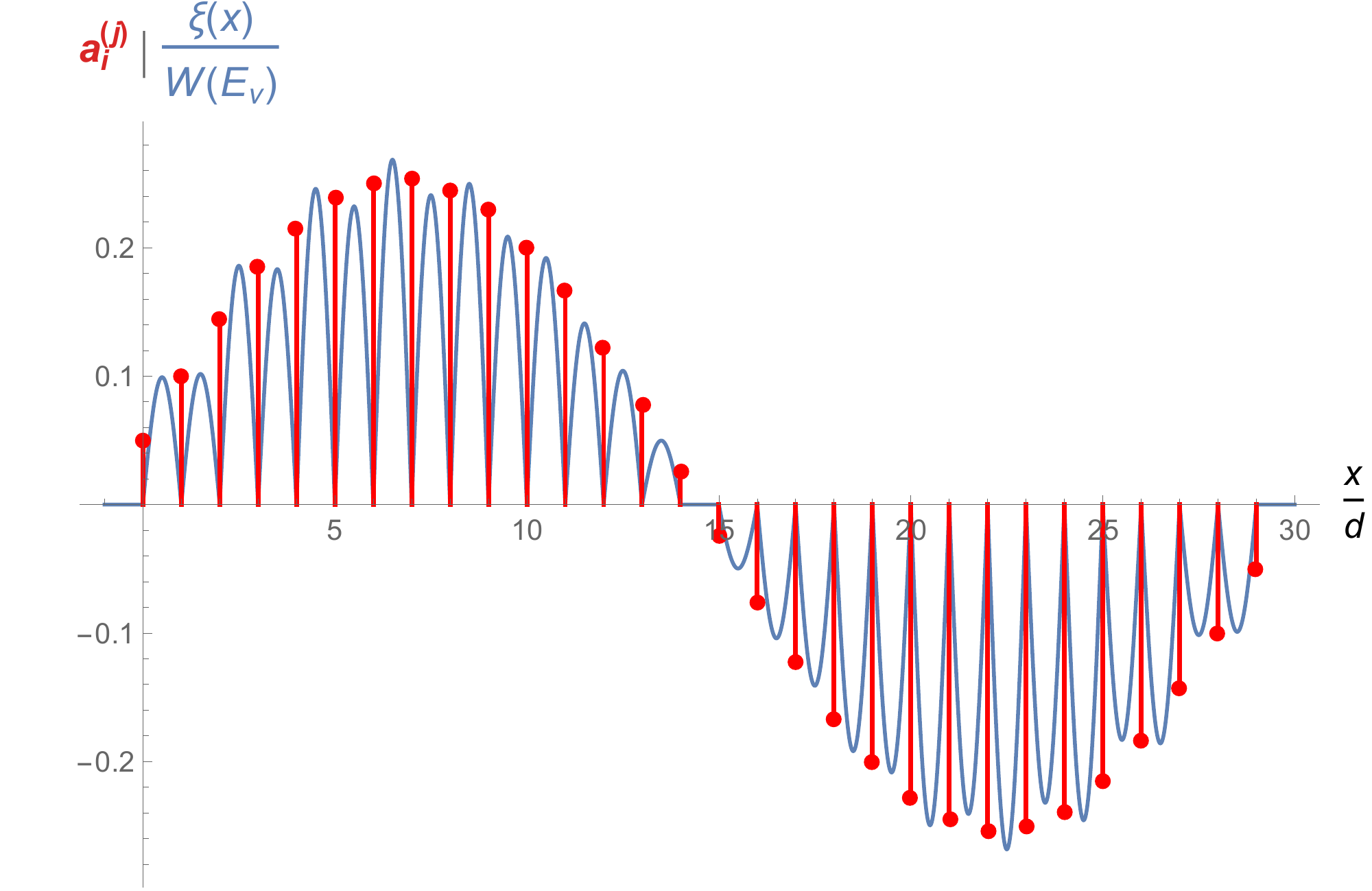}
		\caption{$j=2$}
	\end{subfigure}
	\begin{subfigure}[t]{0.5\linewidth}
		\includegraphics[scale=0.4]{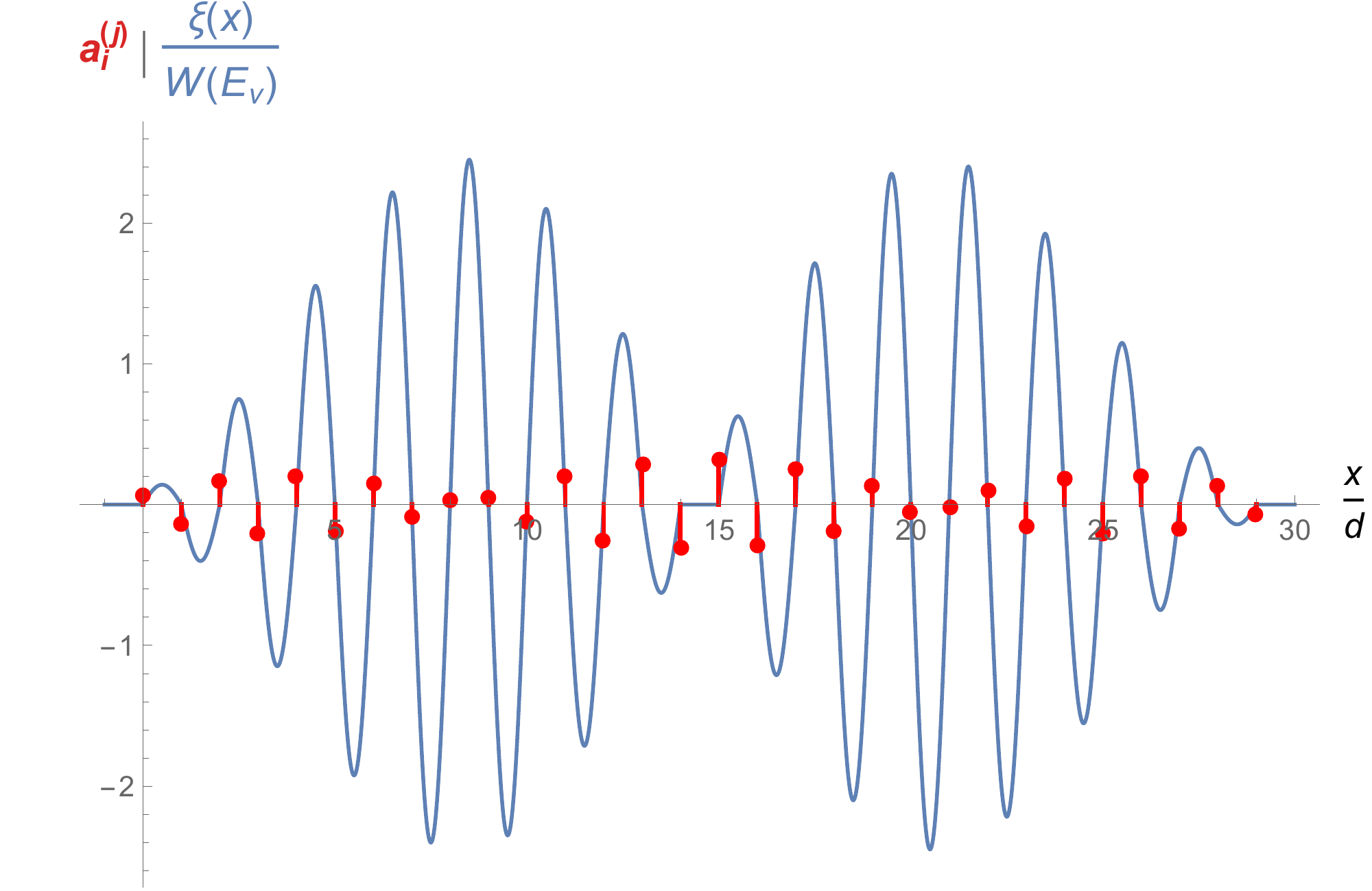}
		\caption{$j=28$}
	\end{subfigure}%
	\begin{subfigure}[t]{0.5\linewidth}
		\includegraphics[scale=0.4]{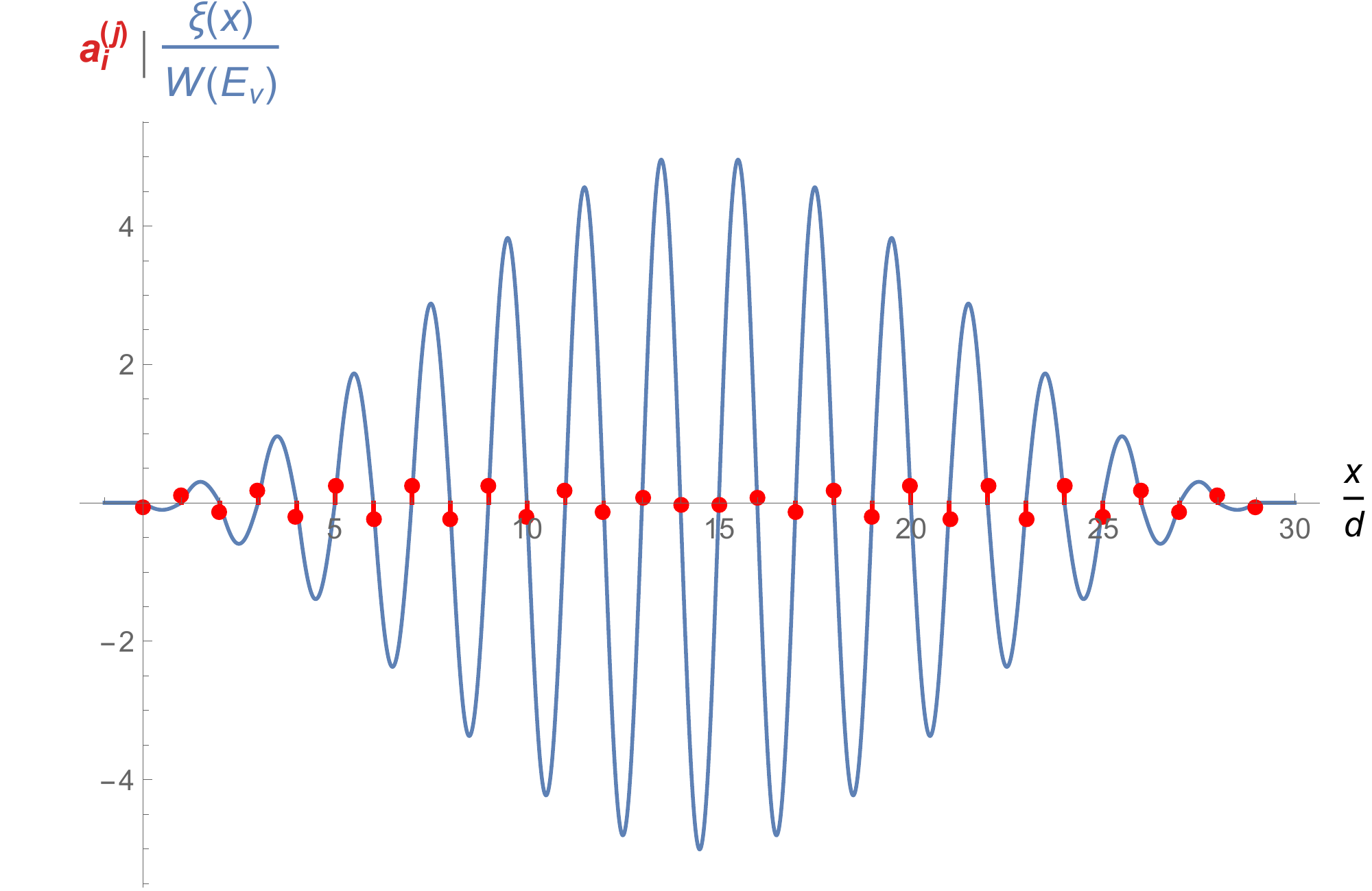}
		\caption{$j=29$}
	\end{subfigure}
	\captionsetup{justification=justified, singlelinecheck=false}
	\caption{Acoustic (upper panels) and optical (lower panels) excitation amplitudes for a system of $n=30$ emitters in BICs with $E=E_1$. Red points at the end of red bars represent the atomic excitation amplitudes $a^{(j)}_\ell$, while the solid blue curve represents the dimensionless field excitation amplitude $\xi(x)/W(E_1)$, with $W(E_1)$ as in Eq.~\eqref{eq:we}. Both quantities reported on the vertical axis are expressed in units of $||\bm{a}^{(j)}||$. The value of each amplitude $a^{(j)}_\ell$ is related to the discontinuity of the first derivative of the field $\xi(x)$ at the position of the $\ell$th emitter, so that each excited emitter induces a jump in the derivative of the photon wavefunction.	
	}
	\label{fig:nuodd}
\end{figure*}

\begin{figure*}[h]
	\centering
	\begin{subfigure}[t]{0.5\linewidth}
		\includegraphics[scale=0.4]{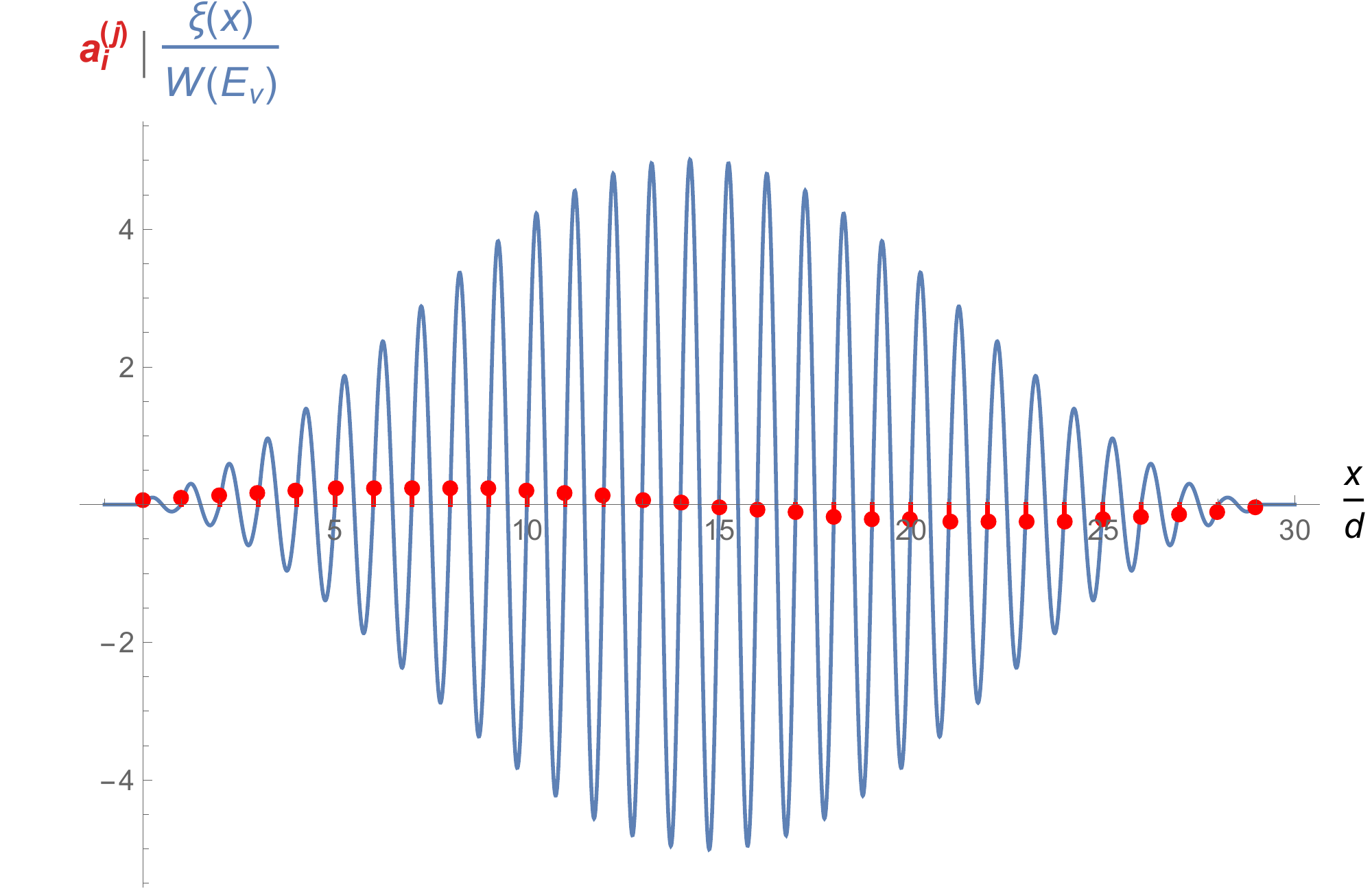}
		\caption{$j=2$}
	\end{subfigure}%
	\begin{subfigure}[t]{0.5\linewidth}
		\includegraphics[scale=0.4]{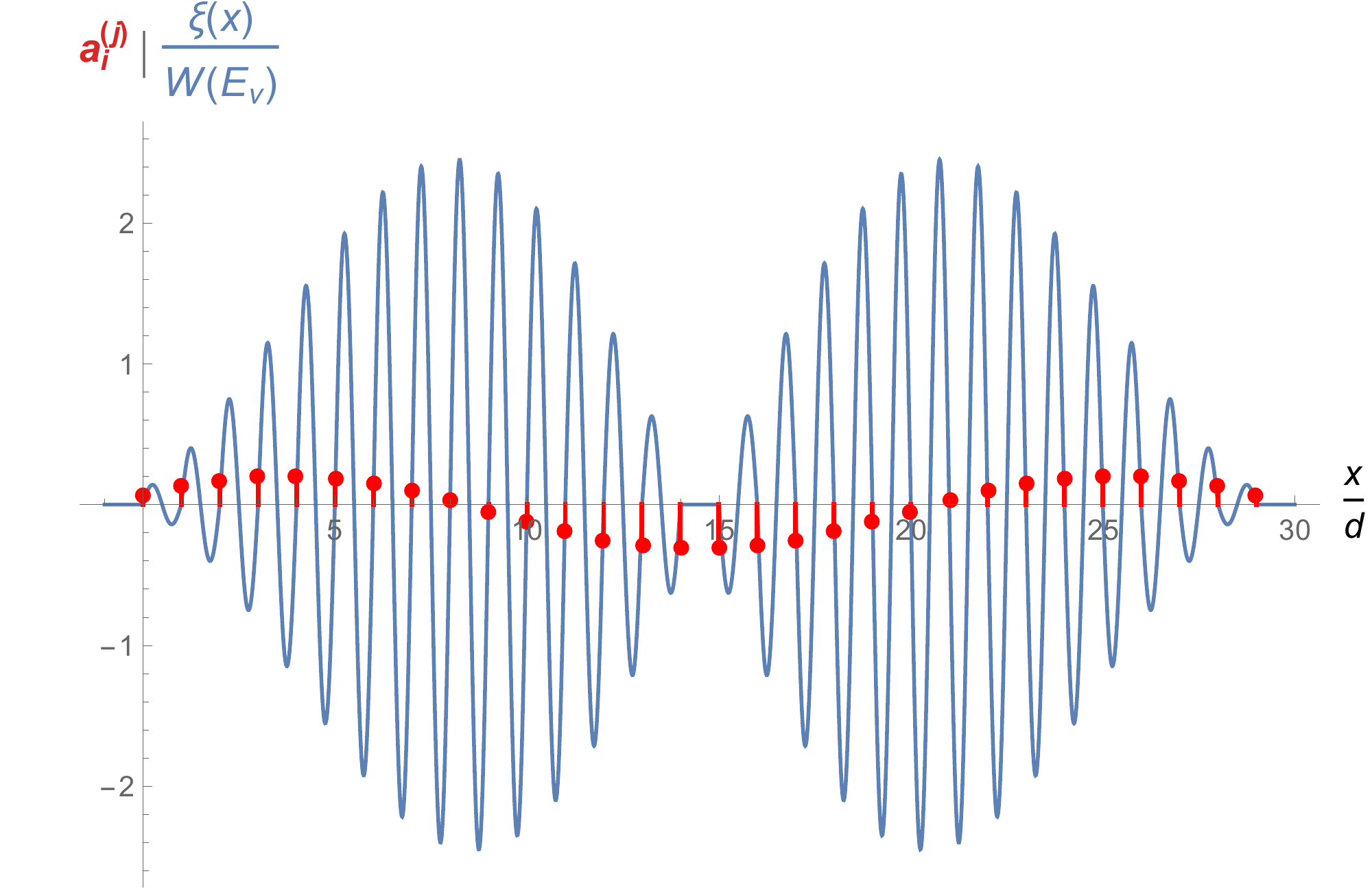}
		\caption{$j=3$}
	\end{subfigure}
	\begin{subfigure}[t]{0.5\linewidth}
		\includegraphics[scale=0.4]{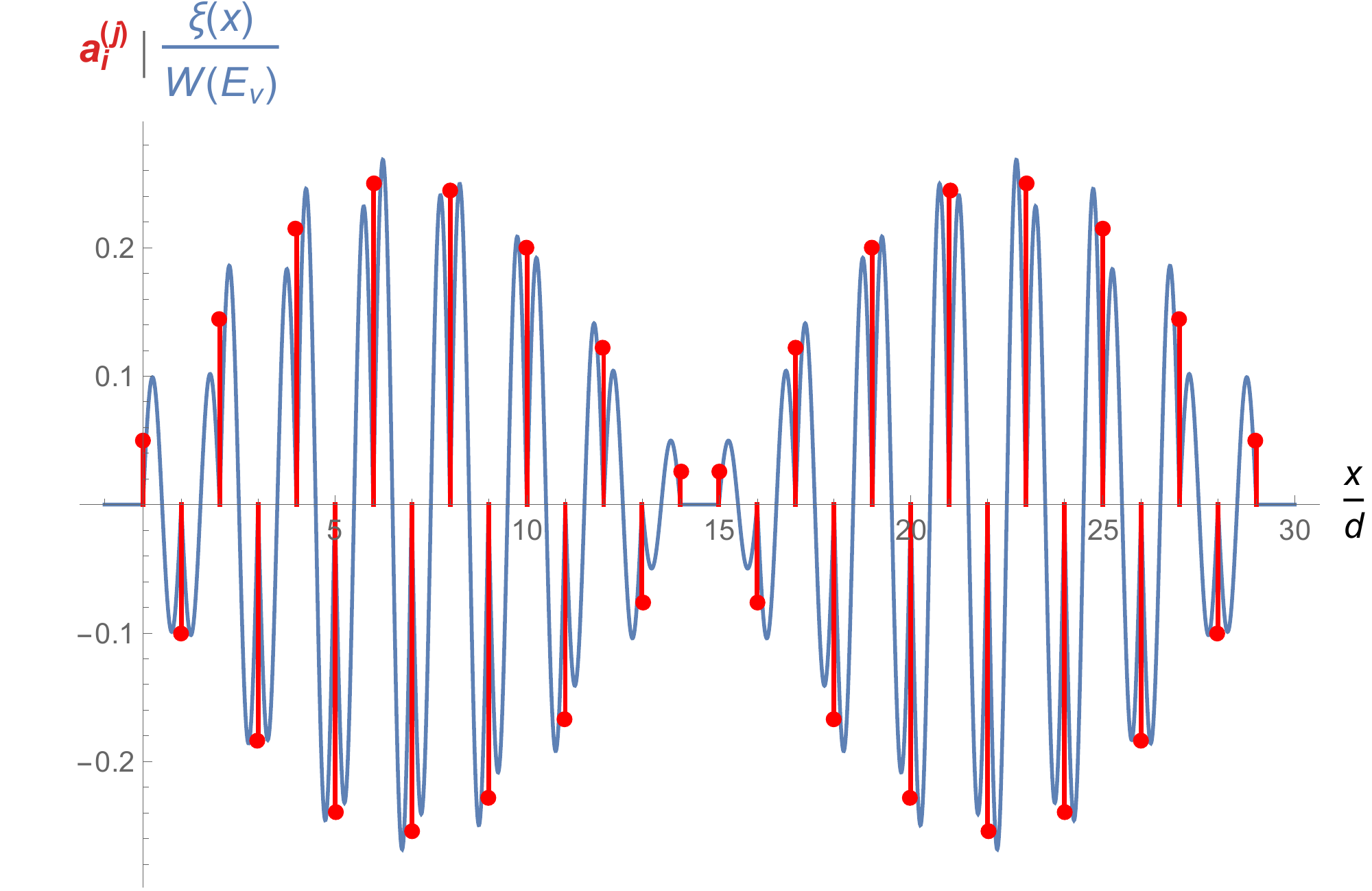}
		\caption{$j=29$}
	\end{subfigure}%
	\begin{subfigure}[t]{0.5\linewidth}
		\includegraphics[scale=0.4]{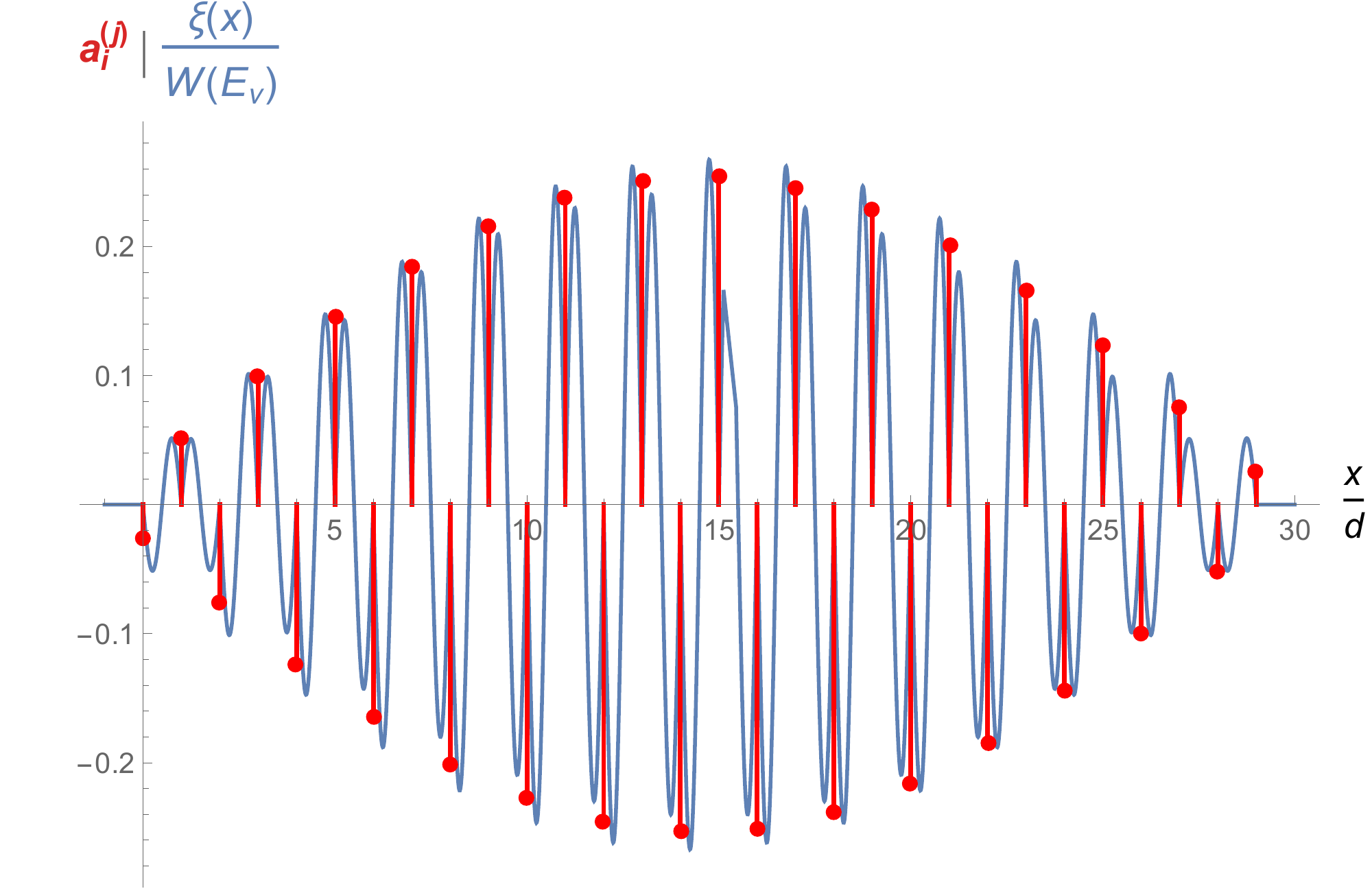}
		\caption{$j=30$}
	\end{subfigure}
\captionsetup{justification=justified, singlelinecheck=false}
	\caption{Acoustic (upper panels) and optical (lower panels) excitation amplitudes for a system of $n=30$ emitters in BICs with $E=E_2$. Red points at the end of red bars represent the atomic excitation amplitudes $a^{(j)}_\ell$, while the solid blue curve represents the dimensionless field excitation amplitude $\xi(x)/W(E_2)$, with $W(E_2)$ as in Eq.~\eqref{eq:we}. Both quantities reported on the vertical axis are expressed in units of $||\bm{a}^{(j)}||$. The value of each amplitude $a^{(j)}_\ell$ is related to the discontinuity of the first derivative of the field $\xi(x)$ at the position of the $\ell$th emitter, so that each excited emitter induces a jump in the derivative of the photon wavefunction.	
	}
	\label{fig:nueven}
\end{figure*}

The BICs of the model are then characterized as follows:
\begin{itemize}
	\item $\lfloor\frac{n}{2}\rfloor$ BICs can emerge at an energy $E=E_\nu$: each of them corresponds \textit{exactly} to one of the sinudoidal waves $\bm{a}^{(j)}$, with $j$ ranging on either even or odd integers as described in Table~\ref{tab:summ}. Each of these BICs is ``switched on'' when the excitation energy $\varepsilon$ of the emitters has \textit{exactly} the value
\begin{equation}\label{eq:varepsilon2}
\varepsilon=E_\nu - \left(\beta_0(E_\nu)+\beta_1(E_\nu)\chi^{(j)}\right).
\end{equation}
	\item $\lceil\frac{n}{2}\rceil$ BICs can emerge at an energy $E\approx E_\nu$: the corresponding excitation profile is a ``deformed" excitation wave $\tilde{\bm{a}}^{(j)}=\bm{a}^{(j)}+\bm{\delta}^{(j)}$, with $j$ ranging on remaining integers, the deformation to be evaluated numerically (see Figs.~\ref{fig:deform}--\ref{fig:deform2} in Appendix~\ref{subsec:nondegen}). Each of these BICs is ``switched on'' when the excitation energy $\varepsilon$ of the emitters has a value 
	\begin{equation}\label{eq:varepsilon3}
	\varepsilon\approx E_\nu - \left(\beta_0(E_\nu)+\beta_1(E_\nu)\chi^{(j)}\right),
	\end{equation}
	to be evaluated numerically as well.
\end{itemize}
The first result follows easily from Eqs.~\eqref{eq:propagator2}--\eqref{eq:propagator0}: if the excitation wave $\bm{a}^{(j)}$ satisfies $\bm{u}_\nu\cdot\bm{a}^{(j)}=0$, then necessarily $\mathrm{G}^{-1}(E_\nu)\bm{a}^{(j)}=0$ provided that Eq.~\eqref{eq:varepsilon2} holds. The second result is nontrivial. We refer to Appendix~\ref{subsec:nondegen} for details. 
\begin{table}[h]\centering
	\begin{tabular}{| c | c | c | c | c |}
		\cline{2-5}
		\multicolumn{1}{c|}{}& \multicolumn{2}{c|}{$n$ even} & \multicolumn{2}{c|}{$n$ odd} \\
		\hline
		& {$j$ even} & {$j$ odd}  & {$j$ even} & {$j$ odd} \\
		\hline
		$\nu$ even & $\bm{a}^{(j)}$ & $\tilde{\bm{a}}^{(j)}$ ($j\neq1$)& $\bm{a}^{(j)}$ & $\tilde{\bm{a}}_\nu^{( j)}$ ($j\neq1$)\\
		$\nu$ odd & $\tilde{\bm{a}}_\nu^{( j)}$ ($j\neq n$)& $\bm{a}^{(j)}$ & $\bm{a}^{(j)}$ & $\tilde{\bm{a}}_\nu^{( j)}$ ($j\neq n$) \\
		\hline
	\end{tabular}
	\captionsetup{justification=justified, singlelinecheck=false}
	\caption{Given a system of $n$ emitters, $n-1$ BICs at energy $E=E_\nu$ (exact excitation waves) or $E\approx E_\nu$ (deformed excitation waves) can emerge, each for a specific value of the excitation energy $\varepsilon$ of the emitters. The $j$th BIC, ordered by increasing frequency, will be exact or deformed depending on whether $n$, $\nu$ and $j$ are even or odd. Notice that either the first ($j=1$) or the last ($j=n$) sinusoidal excitation wave, for $\nu$ even or odd, respectively, will not have any deformed counterpart as eigenvector of the propagator: the corresponding eigenvector will be an unstable state whose eigenvalue has a large imaginary part, i.e.\ a superradiant state.}
	\label{tab:summ}
\end{table}

The physical situation can be described as follows. In order to steadily sustain a BIC, the $n$ emitters must collectively share a part of the excitation following any of the sinusoidal amplitude waves with wavelength~\eqref{eq:lambda}: depending on the values of $\nu$ and $n$, waves with a given parity (either even or odd) will be exact, and waves with the converse parity will be distorted in order to support a stationary excitation of the photon field. 

\subsection{Discussion of the results}\label{subsec:results}

Once we have computed the amplitude waves that can sustain a BIC at energies (close to) $E_\nu$, Eqs.~\eqref{eq:wav}--\eqref{eq:singleboson} give directly the associated photon wavefunction. Figures~\ref{fig:nuodd} and~\ref{fig:nueven} display, for $n=30$, and $E=E_1$ and $E=E_2$, respectively, the simplest acoustic and optical atomic excitation waves that emerge in the emitter array, together with the associated field wavefunction $\xi(x)$, in properly normalized units. Under our assumptions, the photon wavefunction~\eqref{eq:eta} is largely dominated by the contribution of the single pole at $k(E)$. By neglecting the small $\eta$ term, it reads, in correspondence of the $E=E_\nu$ resonance,
\begin{equation}\label{eq:xiresonance}
\xi^{(j)}(x)\simeq  W(E_\nu)\sum_{\ell=1}^n  a^{(j)}_\ell\sin\bigl(\nu\pi|x-(\ell-1)d|\bigr),
\end{equation}
where
\begin{equation}\label{eq:we}
	W(E_\nu)=\sqrt{2\pi} \frac{F(k(E_\nu))}{\omega'(k(E_\nu))}
\end{equation}

It is immediate to see that the contribution \eqref{eq:xiresonance} to the photon amplitude vanishes identically outside the emitter chain. The value of each $a^{(j)}_\ell$ is related to the discontinuity of the first derivative of $\xi(x)$ at the position of the $\ell$th emitter; in particular, the wavefunction is smooth in a neighborhood of the $\ell$th emitter if and only if $a^{(j)}_\ell=0$. In practice, each excited emitter induces a jump, proportional to the excitation amplitude, in the derivative of the photon wavefunction. 

All these results are compatible with the ones for $n=3,4$ reported in~\cite{bic}.

\section{Multimerization}\label{sec:multimerization}
\begin{figure*}[h]
	\centering
	\begin{subfigure}[t]{0.5\linewidth}
		\includegraphics[scale=0.37]{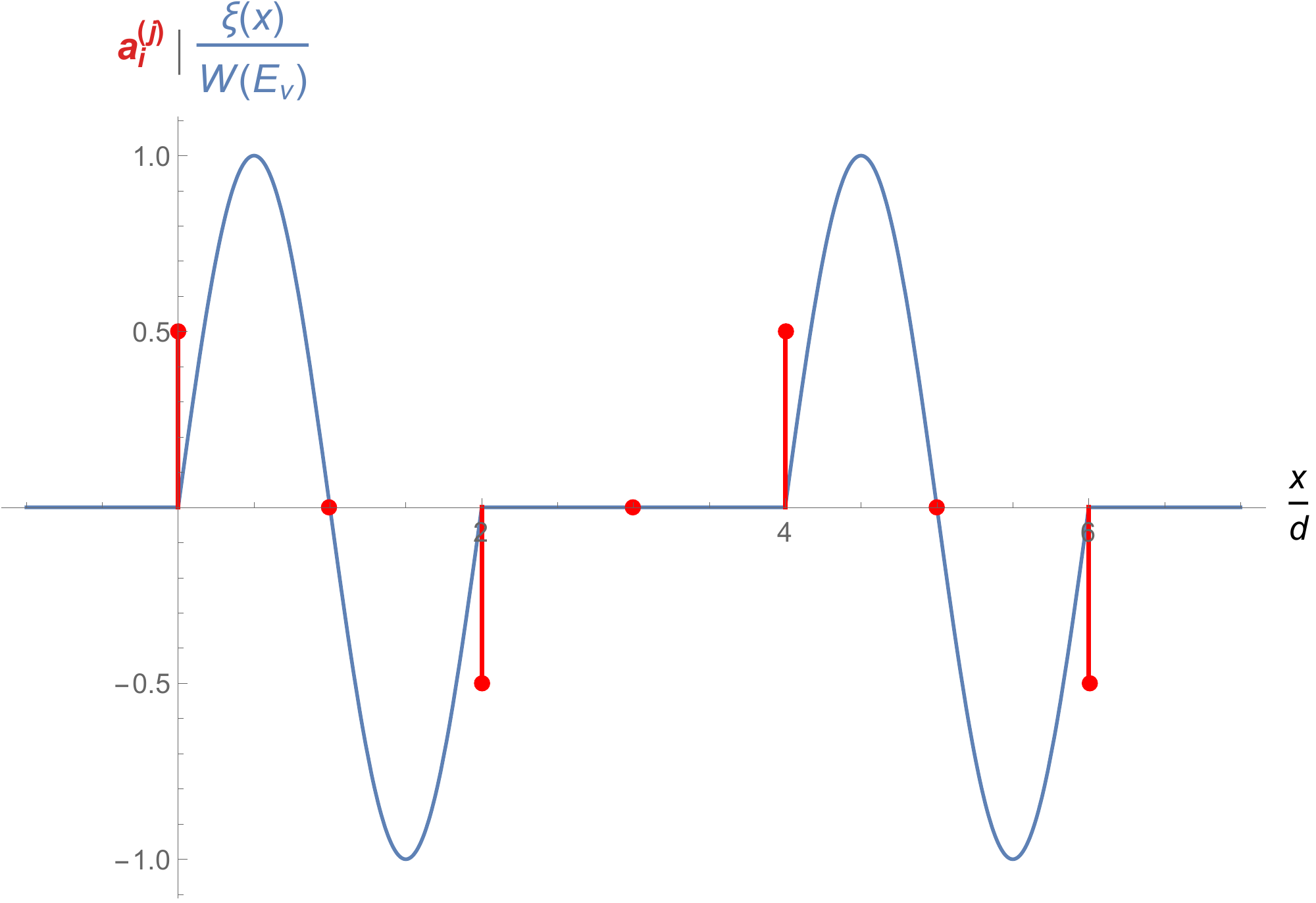}
		\caption{ $n=7$, $r=2$, $h=3$, $j=2$ }
	\end{subfigure}%
	\begin{subfigure}[t]{0.5\linewidth}
		\includegraphics[scale=0.37]{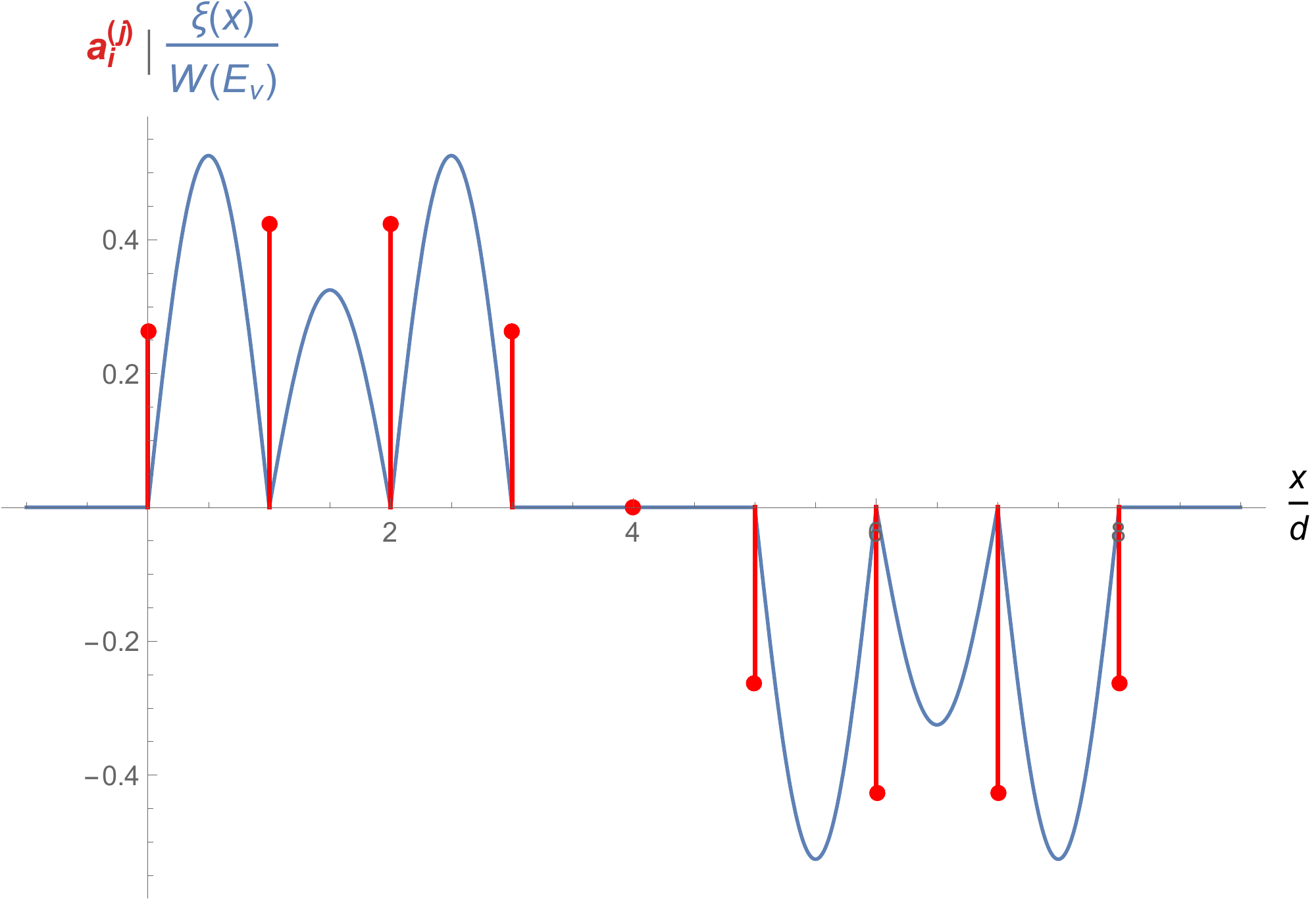}
		\caption{ $n=9$, $r=2$, $h=4$, $j=1$ }
	\end{subfigure}
	\begin{subfigure}[t]{0.5\linewidth}
		\includegraphics[scale=0.37]{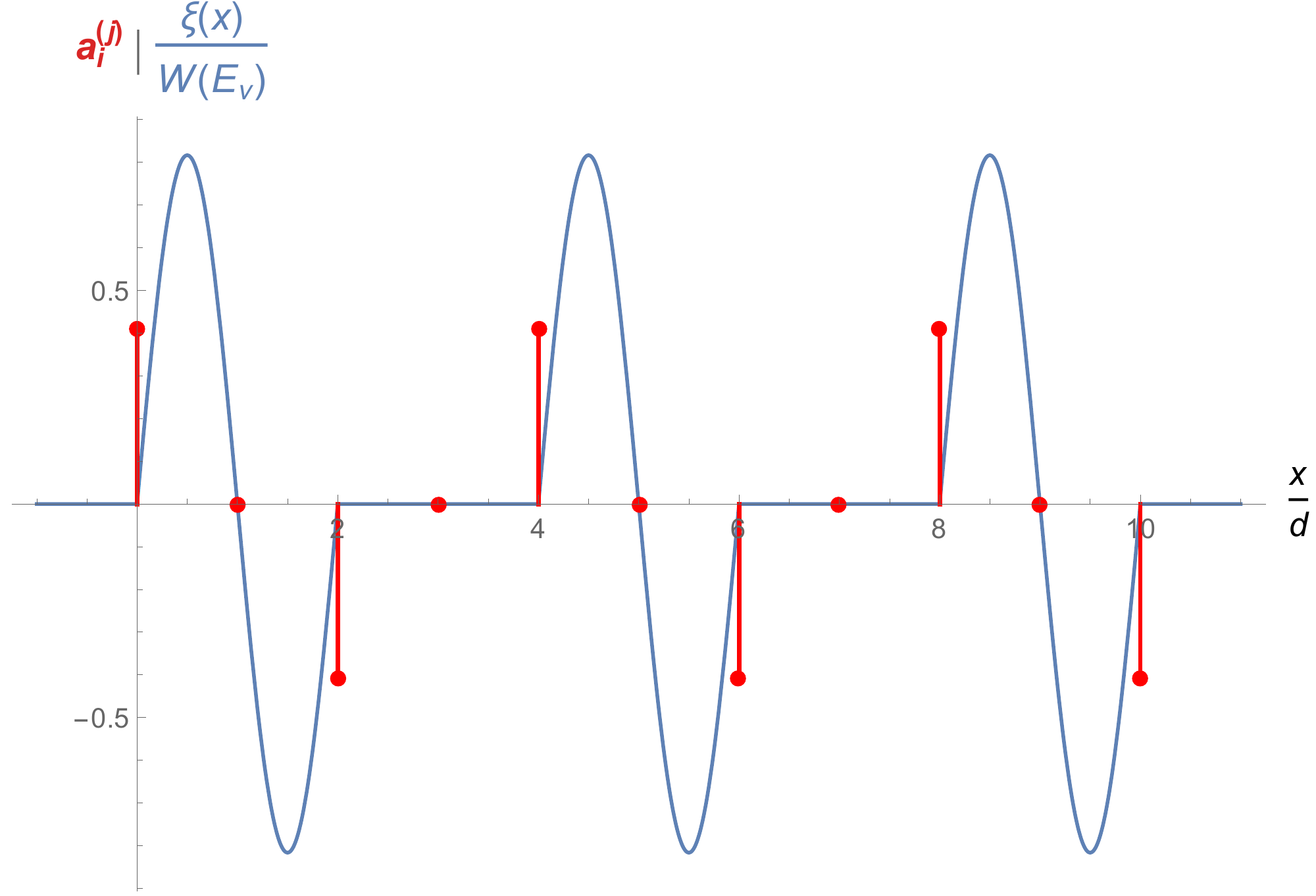}
		\caption{ $n=11$, $r=3$, $h=3$, $j=3$}
	\end{subfigure}%
	\begin{subfigure}[t]{0.5\linewidth}
 		\includegraphics[scale=0.37]{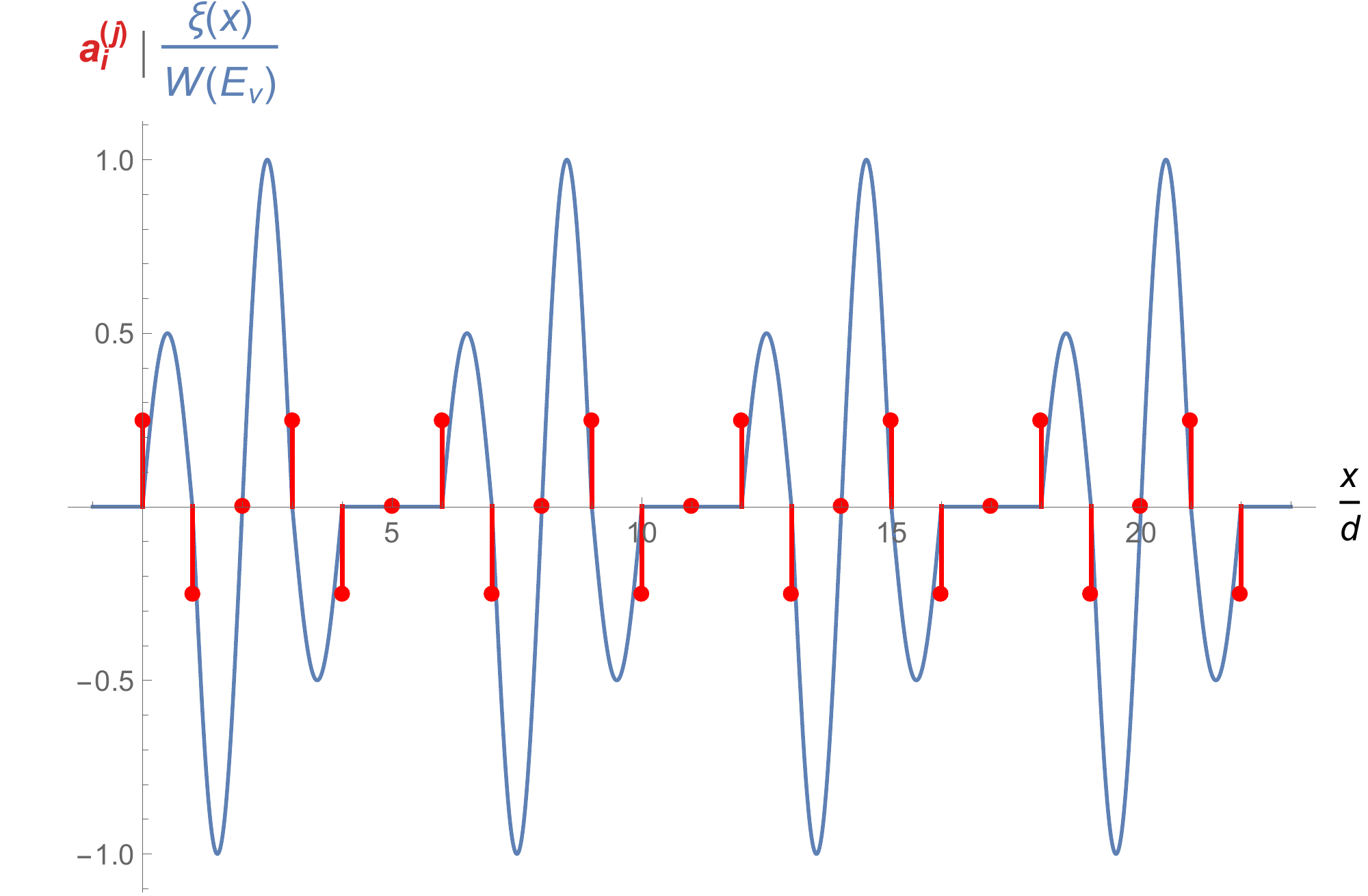}
 		\caption{ $n=23$, $r=4$, $h=5$, $j=8$}
 	\end{subfigure}
\captionsetup{justification=justified, singlelinecheck=false}
	\caption{Examples of exactly resonant multimerized states with energy $E=E_1$: for certain emitter configurations, the photon wavefunction splits in two or more identical waves  separated by  emitters in the ground state. The number of monomers $r$, the number of emitters $h$ involved in a module, and the total number of emitters $n$, are related by $n=rh+r-1$. Both quantities reported on the vertical axis are expressed in units of $||\bm{a}^{(j)}||$.}
	\label{fig:multimer}
\end{figure*}

This final section will be devoted to a better understanding of an interesting phenomenon, which emerges in the structure of the BICs computed in Section~\ref{sec:evaluation}. A detailed scrutiny of the results brings to light a \textit{multimerization} effect: some BICs are characterized by a modular structure, in which the same excitation amplitude configuration repeats for a certain number of times along the chain, with each ``monomer'' separated from the adjacent one by an emitter in its ground state. In the upper panels of Fig.~\ref{fig:multimer}, the configuration is a \textit{dimer}, with the central emitter working as a dynamical mirror \cite{PRA2016}, by forcing the field excitation amplitude to vanish at its position.

This phenomenon is more general: BICs can, in fact, split into more than two parts. A trimer and a tetramer are shown in the lower panels of Fig.~\ref{fig:multimer}. All multimerized states appear to be composed of a number $r$ of identical monomers, made up of $h$ emitters, separated  by a single emitter in its ground state. The number of components and the size of each component are related with $n$ by
\begin{equation}
n= rh + r - 1 .
\end{equation}
Moreover, each module is in itself a BIC for an array of $h$ emitters.

We will interpret this interesting phenomenon as a direct consequence of a simple mathematical property of the adjacency matrix $\Delta_h$ in~\eqref{eq:laplace}, and thus of the particular structure \eqref{eq:propagator2} of the propagator of the model in the nearest-neighbor approximation.

\subsection{Structure of multimers}
Consider an integer $h$, and define two column vectors
\begin{equation}
\bm{x}_h= (0, 0, \dots, 0, 1)^\intercal, \qquad
\bm{y}_h = (1, 0, \dots, 0, 0)^\intercal.
\end{equation}
Then, given $n,r\in\mathbb{N}$ with $n=rh+(r-1)$, and decomposing the space $\mathbb{C}^n$ as
\begin{equation}
\mathbb{C}^n=\overbrace{
	(\mathbb{C}^h\oplus\mathbb{C})\oplus(\mathbb{C}^h\oplus\mathbb{C})\oplus\cdots(\mathbb{C}^h\oplus\mathbb{C})
}^
{r\text{ times}}
\oplus\,\mathbb{C}^h,
\end{equation}
i.e., representing vectors $\bm{A}\in\mathbb{C}^n$ as
\begin{equation}\renewcommand{\arraystretch}{1.3}\label{eq:arrayblock}
\bm{A}= (\bm{a}^{(1)}, b^{(1)}, \bm{a}^{(2)}, b^{(2)},
\dots, \bm{a}^{(r-1)}, b^{(r-1)}, \bm{a}^{(r)})^\intercal,
\end{equation}
with $\bm{a}^{(1)},\dots,\bm{a}^{(r)}\in\mathbb{C}^h$ and $b^{(1)},\dots,b^{(r-1)}\in\mathbb{C}$, the matrix $\Delta_n$ with dimension $n=rh+(r-1)$ takes the following block structure
\begin{equation}\label{eq:laplaceblock}\renewcommand{\arraystretch}{1.3}
\Delta_{n}=\begin{pmatrix}
\Delta_h&\bm{x}_h&\\
\bm{x}_h^\intercal &0&\bm{y}_h^\intercal \\
&\bm{y}_h &\Delta_h&\bm{x}_h&\\
&&\bm{x}_h^\intercal &0&\bm{y}_h^\intercal \\
&&&\bm{y}_h &\Delta_h&\bm{x}_h&\\
&&&&\ddots&\ddots&\ddots\\
&&&&&\bm{x}_h^\intercal &0&\bm{y}_h^\intercal \\
&&&&&&\bm{y}_h &\Delta_h
\end{pmatrix},
\end{equation}
with the rows of odd order acting onto the ``vector'' components $\bm{a}^{(s)}$ in~\eqref{eq:arrayblock} and the rows of even order acting onto the ``scalar'' ones $b^{(s)}$. In particular, if we consider a vector as in~\eqref{eq:arrayblock} with all scalar components being zero, i.e. 
\begin{equation}\renewcommand{\arraystretch}{1.0}\label{eq:arrayblock2}
\bm{A}= (\bm{a}^{(1)}, 0, \bm{a}^{(2)}, 0, 
\dots, \bm{a}^{(r-1)}, 0, \bm{a}^{(r)})^\intercal ,
\end{equation}
one gets

\begin{equation}\renewcommand{\arraystretch}{1.3}\label{eq:lapl}
\Delta_n\bm{A}= \bigl(\Delta_h\bm{a}^{(1)},\, a^{(1)}_h+a^{(2)}_1,\, \Delta_h\bm{a}^{(2)},\, a^{(2)}_h+a^{(3)}_1,
\dots, a^{(r-1)}_h+a^{(r)}_1,\, \Delta_h\bm{a}^{(r)} \bigr)^\intercal.
\end{equation}
Therefore, an array characterized by $a^{(s)}_h= - a^{(s+1)}_1$ for all $s=1,\dots,r$ (i.e.\ with the first component of the $s$th ``block'' being the opposite of the last component of the $(s-1)$th block), satisfies
\begin{equation}
\Delta_n\bm{A}= \bigl( \Delta_h\bm{a}^{(1)},\, 0,\, \Delta_h\bm{a}^{(2)},\, 0,
\dots, 0,\, \Delta_h\bm{a}^{(r)} \bigr)^\intercal.
\end{equation}

In such a case, the action of the $n$-dimensional matrix $\Delta_n$ on the full array splits into the action of the $h$-dimensional matrix $\Delta_h$ on each block. As a direct consequence:
\begin{itemize}
	\item if $\bm{a} \in\mathbb{C}^h$ is an \textit{antisymmetric} eigenvector of $\Delta_h$, then $a_h=-a_1$ and the vector
	\begin{equation}\label{eq:multimer}
	\bm{A}=(\bm{a}, 0,\bm{a},0,\bm{a},0,\dots)^\intercal
	\end{equation}
	is an eigenvector of $\Delta_n$ with the same eigenvalue;
	\item if $\bm{a} \in\mathbb{C}^h$ is a \textit{symmetric} eigenvector of $\Delta_h$, then $a_h=a_1$ and the vector
	\begin{equation}\label{eq:multimer2}
	\bm{A}=(+\bm{a},0,-\bm{a},0,+\bm{a},0,\dots)^\intercal
	\end{equation}
is an eigenvector of $\Delta_n$ with the same eigenvalue.
\end{itemize}

It is easy to show that the mathematical property outlined above, together with the expression~\eqref{eq:propagator2} for the propagator of the system in the nearest-neighbor approximation, does explain the multimerization phenomenon outlined at the start of this section. As discussed in Subsection~\ref{subsec:deform} (also see Table~\ref{tab:summ}), an $h$-emitter excitation wave $\bm{a}$ (i.e. an eigenvector of $\Delta_h$) with the ``right'' parity, i.e. such that $\bm{u}_\nu\cdot\bm{a}=0$, satisfies
\begin{equation}
	\mathrm{G}^{-1}(E_\nu)\bm{a}=0
\end{equation}
provided Eq.~\eqref{eq:varepsilon} holds, and is thus associated with a BIC for a chain of $h$ identical emitters. Now, the $n$-component vector $\bm{A}$ defined either via Eq.~\eqref{eq:multimer} or~\eqref{eq:multimer2} (depending on the value of $\nu$) is an eigenvector of $\Delta_n$ with the same eigenvalue \textit{and}, by construction, satisfies the condition $\bm{U}_\nu\cdot\bm{A}=0$, with $\bm{U}_\nu$ being the $n$-component analogue of $\bm{u}_\nu$: consequently, it is an eigenstate of the propagator and thus corresponds to a multimerized BIC for the $n$-emitter chain.

\subsection{Discussion of the results}
Our findings demonstrate that it is always possible to construct a multimerized BIC of $n=rh+(r-1)$ emitters from smaller (building) blocks of $h$ emitters, separated by single emitters that act as dynamical mirrors. The rules for constructing such states are, nevertheless, very specific: the blocks must be equal and the way of connecting them to each other must follow one of the prescriptions \eqref{eq:multimer}-\eqref{eq:multimer2}, depending on their symmetry. These conditions are ultimately due to the small corrections to the propagator and the wavefunctions due to the evanescent fields. 

In the absence of evanescent fields, multimerized configurations could be constructed arbitrarily among the infinite possibilities in a degenerate $(n-1)$-dimensional eigenspace, by superposing ``local'' excitations involving few neighboring emitters: this is due to the fact that adjacent monomers would not interact with each other, since no photon field would be present between them, and our model only involves photon-mediated interactions. On the other hand, the presence of evanescent fields modifies the picture by inducing interactions between adjacent monomers, thus making most of multimerized one-excitation states unstable, and selecting only a few ones as stable configurations. Hence, the role of the evanescent fields is to induce the emergence of collective stable states. In particular, our findings show that one emitter in its ground state must be present between neighboring monomers in order to ensure stability: in this sense, even the emitter that does not share the excitation cooperates with the rest of the system to form the bound state.

\section*{Conclusions and outlook}

We analyzed the emergence of BICs in a regular array of quantum emitters in a waveguide. BICs are present for an arbitrary number of emitters, and the excitation profile of the emitter states is a sinusoidal wave. We also discussed the presence of multimers, separated by two regions in which there is practically no electromagnetic field.
A crucial role is played by the evanescent fields generated by quantum emitters in determining the physical structure of bound states in the continuum.

The techniques adopted in this article hinge upon an analysis of the propagator and are intrinsically non-perturbative. The main physical factor that limits the robustness of BICs is photon loss from the waveguide: we did not discuss here these losses, but their effect can be estimated by applying standard techniques based on the master equation.

The collective behaviour unearthed in our analysis does not depend on the specific geometry of the waveguide and the details of the dispersion relations, and is therefore valid for a number of physical implementations. The identification of (manifolds of) states in which a single excitation is coherently shared among distant artificial atoms enables one to analyze the features of long-range coherence, mediated by the photon field, and could possibly provide a tool to control qubits at an arbitrary distance. 

Moreover, the manifolds of long-lived states, whose lifetime is determined by the waveguide losses and not by the atom-field coupling, can be used as quantum registers/memories \cite{register} due to their dynamical stability, in particular against spatially separated decoherence sources~\cite{memo}. These features can make them useful in hydrid situations~\cite{hyb,hyb2}. 

Finally, we mention that additional properties emerge when the emitters are confined in a \emph{finite} waveguide \cite{ringWQED} or in 2-dimensional geometries \cite{feiguin}, or when topological effects contribute to the robustness of the dressed states \cite{leonforte}.
These situations and related aspects will be investigated in the future.

\section*{Acknowledgments}
PF and SP acknowledge support by MIUR via PRIN 2017 (Progetto di Ricerca di Interesse Nazionale), project QUSHIP (2017SRNBRK). 
PF and DL are partially supported by the Italian National Group of Mathematical Physics (GNFM-INdAM). 
 PF, DL, SP, and FVP are partially supported by Istituto Nazionale di Fisica Nucleare (INFN) through the project ``QUANTUM'' and by Regione Puglia and  QuantERA ERA-NET Cofund in Quantum Technologies (GA No.\ 731473), project PACE-IN\@.

\appendix

\section{Eigenvalue equation}

\label{sec:eigenvalue}
In this appendix we derive the eigenvalue equations for BICs discussed in Sec.~\ref{sec:model}.
Consider the eigenvalue equation 
\begin{equation}
(H - E) \ket{\Psi} =0,
\end{equation}
where $H$ is Hamiltonian~\eqref{tothamiltonian}--\eqref{Hint} and $\ket{\Psi}$ state~\eqref{state1}, and project it onto the basis vectors $\ket{e_j} \otimes \ket{\mathrm{vac}}$ and  $\ket{g} \otimes b^{\dagger}(k) \ket{\mathrm{vac}}$. One gets 
\begin{align}
&(\varepsilon-E) a_j + \int \dd k\,   F(k)\e^{\ii (j-1)kd}  \,\tilde{\xi}(k) =0, \nonumber\\
&(\omega(k)-E) \, \tilde{\xi}(k) + \sum_\ell  F(k)\e^{-\ii (\ell-1)kd} a_\ell =0.
\label{eq:A2}
\end{align}
The second equation gives
\begin{equation}
\tilde{\xi}(k) = \sum_\ell \frac{F(k)}{E - \omega(k)}  \e^{-\ii (\ell-1)kd}a_\ell, \quad \text{if } \omega(k)\neq E,
\label{eq:A3}
\end{equation}
and
\begin{equation}
\sum_\ell F(k)\e^{-\ii (\ell-1)kd} a_\ell = 0, \quad \text{if } \omega(k)=E;
\end{equation}
these yield, respectively, Eq.~\eqref{eq:wav} by a Fourier transform, and the constraint~\eqref{eq-physicalcondition}, which is needed in order to make the integral in Eq.~\eqref{eq:wav} well-defined.

 By plugging~\eqref{eq:A3} into the first equation in \eqref{eq:A2} we get
\begin{equation}
(\varepsilon-E) a_j +  \int \dd k\,  \sum_\ell \frac{|F(k)|^2}{E-\omega(k)} \e^{\ii(j-\ell)kd}\, a_\ell=0, 
\end{equation}
which, by defining $\mathrm{G}^{-1}(E)$ as in Eq.~\eqref{eq:propagator}, is Eq.~\eqref{eq:BICs2}. The latter equation admits nontrivial solutions if and only if Eq.~\eqref{eq:BICs1} holds.

\section{Calculation of the propagator}\label{app:self}
In this appendix we will compute the propagator for our model, thus showing the second part of Eq.~\eqref{eq:propagator}. Let us introduce the self-energy matrix:
\begin{equation}\label{eq-selfenergy}
\Sigma_{j\ell}(z) = \int \dd k\,  \frac{|F(k)|^2}{z-\omega(k)}\e^{\ii(j-\ell)kd},\qquad \Im z>0,
\end{equation}
so that the propagator can now be expressed as
\begin{equation}
	\mathrm{G}^{-1}(E)=(\varepsilon-E)\mathrm{I}_n+\Sigma(E+\ii0).
\end{equation}
We will find an expression for the self-energy matrix~\eqref{eq-selfenergy}, and thus for the propagator, under general assumptions on the dispersion relation
$\omega(k)$ and the coupling function $F(k)$.

We make the following assumptions: 
\begin{itemize}
\item [(i)] $\omega(k)$ and $f(k)=|F(k)|^2$ are even real-valued functions satisfying the integrability condition:
\begin{equation}
\int_{\mathbb{R}} \frac{f(k)}{\omega(k)+1}\,\dd k<\infty;
\end{equation}

\item [(ii)] $\omega'(k)>0$ for all $k>0$, so that $\omega(k)$ is an increasing function for $k>0$, and $\omega(0)>0$.

\item [(iii)] $\omega(k)$ and $f(k)$ have complex analytic continuations $\omega(\kappa)$, $f(\kappa)$ in a strip $\mathcal{K}=\{k+ \ii \eta\, : \, k \in \mathbb{R}, \eta \in (-m,m)\}$ of the complex plane 
and are continuous up to $\partial\mathcal{K}= (\mathbb{R} + \ii m) \cup (\mathbb{R} - \ii m)$;

\item [(iv)] the following conditions hold: 
\begin{equation}
\lim_{ |\Re \kappa | \to  \infty}\, \frac{f( \kappa)}{\omega( \kappa)}=0,
\end{equation}
uniformly in the strip $\mathcal{K}$, and
\begin{equation}
\int_{\mathbb{R}}\frac{|f(k+\ii m)|}{|\omega(k+\ii m)|+1}\,\mathrm{d}k <\infty;
\label{eq:intbound}
\end{equation}

\item [(v)] there is an open connected set $\mathcal{A}\subset\mathbb{C}$ containing the real half line $(\omega(0), +\infty)$ such that, for $z\in\mathcal{A}$, the equation $\omega(\kappa)=z$ admits exactly two solutions in $\mathcal{K}$;

\end{itemize}
We remark that an expression for the self-energy can be found without condition (iii),  and that the discussion that follows can be extended to a nonmonotonic $\omega(k)$, provided that one looks for the eigenvalues in an energy range $(E_1,E_2)$ such that the equation $\omega(k)=E$ has a single positive real solution for all $E\in(E_1,E_2)$.

First of all, since both $\omega(k)$ and $f(k)$ are even real-valued functions, i.e. satisfy, for all $k\in\mathbb{R}$,
\begin{align}
\omega(-k)&=\omega(k),\qquad \omega(k)^*=\omega(k);\\
f(-k)&=f(k),\qquad f(k)^*=f(k)
\end{align}
their analytic continuations $\omega(\kappa)$, $f(\kappa)$ to the strip $\mathcal{K}$ of the complex plane, with $\mathcal{K}^\pm=\mathcal{K}\cap\mathbb{C}^\pm$, satisfy the following symmetry properties:
\begin{align}
\omega(-\kappa)&=\omega(\kappa),\qquad \omega(\kappa)^*=\omega(\kappa^*);\\
f(-\kappa)&=f(\kappa),\qquad f(\kappa)^*=f(\kappa^*).
\label{eq:symmp}
\end{align}
The self-energy matrix is defined for $j,\ell=1,\ldots,n$ and $z\in\mathbb{C}$ by
\begin{equation}
\Sigma_{j\ell}(z)=\int_{-\infty}^{\infty}\frac{f(k)}{z-\omega(k)}\e^{\ii (j-\ell)kd}\,\mathrm{d}k,
\end{equation}
and, because of the symmetry properties of the integrand, satisfies
\begin{equation}
\Sigma_{j\ell}(z)=\Sigma_{\ell j}(z)
\end{equation}
for all $z\in\mathbb{C}$ and $j,\ell=1,\dots,n$, so that we only need to compute it for $j\geq\ell$ or, equivalently, we can substitute $j-\ell$ with $|j-\ell|$.

\begin{figure}\centering
	\includegraphics[scale=1.2]{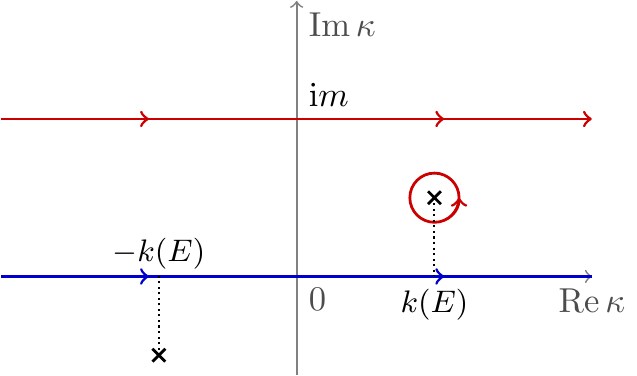}
	\captionsetup{justification=justified, singlelinecheck=false}
	\caption{Representation of the integration contour and the analyticity strip in the  complex $\kappa$ plane. The integral over $\mathbb{R}$ (blue lower line) equals the integral over $\mathbb{R}+\ii m$ (red upper line) plus the residue of the integrand at the pole (red circle).}
	\label{fig-contour}
\end{figure}

We will evaluate the integral 
\begin{equation}
\label{intintegral}
\Sigma_{j\ell}(z)=\lim_{R\to\infty}\int_{-R}^{R}\frac{f(k)}{z-\omega(k)}\e^{\ii |j-\ell|kd}\,\mathrm{d}k
\end{equation}
through a contour integration in the complex plane (see Fig.\ \ref{fig-contour}), by closing the integration over $[-R,R]$ with a properly chosen curve and exploiting the residue theorem.
The integrand in (\ref{intintegral}) has poles at the solutions of the equation $\omega(\kappa)=z$. Let us consider $z\in\mathcal{A}$: since $\omega(-\kappa)=\omega(\kappa)$, by assumption (v) such solutions come in pairs $\pm\kappa(z)$.
By the analyticity of $\omega(\kappa)$, they are locally continuous and, moreover, analytic away from the critical points of $\omega(\kappa)$, i.e.\ the points at which $\omega'(\kappa)=0$. Aside from such (countably many) points, each solution $\kappa(z)$ is a \textit{simple} pole for the integrand.

Choose $R>0$ large enough so that the pair of poles is included in the rectangle
\begin{equation}
\mathcal{K}_R=\left\{\kappa\in\mathcal{K}:\,|\Re \kappa|\leq R\right\},
\end{equation}
and $\mathcal{K}^\pm_R=\mathcal{K}_R\cap\mathbb{C}^\pm$. We will close the integration contour on the boundary of $\mathcal{K}_R^+$, consisting of  two horizontal segments, $[-R, R]$ and  $[-R,R]+\ii m$, and two vertical segments, $\pm R + \ii[0,m]$.
 For $j=\ell$, by hypothesis (iv) the integral on $\partial\mathcal{K}^+$ converges and there is no contribution from the vertical segments as $R\to\infty$, hence
\begin{align}
&\lim_{R\to\infty}\int_{\partial\mathcal{K}^+_R}\frac{f(\kappa)}{z-\omega(\kappa)}\,\mathrm{d}\kappa  =\int_{\partial\mathcal{K}^+}\frac{f(\kappa)}{z-\omega(\kappa)}\,\mathrm{d}\kappa
\nonumber\\
& =\int_{\mathbb{R}}\frac{f(k)}{z-\omega(k)}\,\mathrm{d}k - \int_{\mathbb{R}}\frac{f(k+\ii m)}{z-\omega(k+\ii m)}\,\mathrm{d}k.
\end{align}
For $j\neq\ell$, we have
\begin{equation}
\bigg|\frac{f(\kappa)}{z-\omega(\kappa)}\,\e^{\ii |j-\ell|\kappa d}\bigg|\leq\,\e^{-\operatorname{Im}\kappa|j-\ell|d}
\bigg|\frac{f(\kappa)}{z-\omega(\kappa)}\bigg|,
\end{equation}
and thus, 
\begin{equation}
\bigg|\int_{\mathbb{R}+\ii m}\frac{f(\kappa)}{z-\omega(\kappa)}\,\e^{\ii |j-\ell|\kappa d}\mathrm{d}\kappa\bigg|\leq\,\e^{-|j-\ell| m d}\int_{\mathbb{R}}\bigg| \frac{f(k+\ii m)}{z-\omega(k+\ii m)}\bigg| \mathrm{d}k <\infty,
\end{equation}
i.e. all off-diagonal contributions are finite as well and, moreover, exponentially suppressed in  $m d$.

The integral over the contour will equal the sum of the residua of all singularities enclosed by the contour, that is, all solutions of the equation $\omega(\kappa)=z$ with positive imaginary part. By our assumptions, for $z\in\mathcal{A}$ there is precisely one couple of singularities $\pm\kappa(z)$, with $\Im \kappa(z)>0$ for all $z\in\mathcal{A}\cap\mathbb{C}^+$, each being continuously linked with the corresponding real solution $\pm k(E)$ of the real equation $\omega(k)=E$ for $E>\omega(0)$. Indeed, by taking into account the equation
\begin{equation}
\omega(k+\ii \eta)=E+\ii \delta,
\end{equation}
and expanding around the solution $k(E)$ of the equation $\omega(k)=E$ with $\omega'(k(E))> 0$, we get
\begin{equation}
\eta\sim\frac{\delta}{\omega'(k(E))},
\end{equation}
thus implying that the pole contained in the integration contour is $+\kappa(z)$, as shown in Fig.~\ref{fig-contour}.

By the residue theorem we thus have
\begin{equation}
\int_{\partial\mathcal{K}^+} g_{j\ell}(\kappa) \, \dd \kappa
= \Sigma_{j\ell}(z) - \int_{\mathbb{R}} g_{j\ell}(k+\ii m) \, \dd k
= 2\pi \ii  \, \mathrm{Res} \big( g_{j\ell} ,  \kappa(z)\big),
\end{equation}
where
\begin{equation}
g_{j\ell}(\kappa) = \frac{f(\kappa)}{z-\omega(\kappa)}\, \e^{\ii |j-\ell|\kappa d},
\end{equation}
and $\mathrm{Res} \big( g_{j\ell} ,  \kappa(z)\big)$ denotes its residue at the pole $\kappa(z)$. For all noncritical values of $z$, we have
\begin{equation}
z-\omega(\kappa)\sim - \omega'(\kappa(z))(\kappa-\kappa(z)),\quad \text{as} \quad \kappa\to\kappa(z),
\end{equation}
hence
\begin{equation}
\lim_{\kappa\to\kappa(z)}(\kappa-\kappa(z))\frac{f(\kappa)}{z-\omega(\kappa)}=-\frac{f(\kappa(z))}{\omega'(\kappa(z))}
\end{equation}
and thus the residue  is
\begin{equation}
\mathrm{Res} \big( g_{j\ell} ,  \kappa(z)\big)= -\frac{f(\kappa(z))}{\omega'(\kappa(z))}\e^{\ii |j-\ell|\kappa(z)d}.
\label{eq:res}
\end{equation} 

Thus we finally obtain the desired result: for all $E>\omega(0)$,
\begin{equation}
\Sigma_{j\ell}(E+\ii 0)= -\ii Z(E)\e^{\ii | j-\ell | k(E)d} +  \beta_{j-\ell}(E) ,
\label{eq:B23}
\end{equation}
with a residue term
\begin{equation}\label{eq:ze}
Z(E)= 2 \pi \frac{f(k(E))}{\omega'(k(E))}\geq 0,
\end{equation}
and a contour contribution
\begin{equation}
\beta_{j}(E) = \e^{- | j | m d} \int_{\mathbb{R}}\frac{f(k+\ii m)}{E- \omega(k+\ii m)} \, \e^{\ii |j| k d}\,\dd k.
\label{eq:B25}
\end{equation}
Notice that $\beta_{j}(E)= \beta_{j}(E)^*$ is real, since by using the symmetry properties~\eqref{eq:symmp} it easy to see that the integrand function has the symmetry $g_{j\ell}(k+\ii m)^* = g_{j\ell}(-k+\ii m)$. Therefore, the contour term gives a contribution to the self-energy matrix $\Sigma(E)$ which is  real Hermitian, whereas the residue yields a contribution which is non-Hermitian. 

\section{Calculation of the photon wavefunction}

\label{sec:phwave}

In this appendix we apply the results of Appendix~\ref{app:self} in order to compute the  integral in~\eqref{eq:wav}, which gives  the boson wavefunction $\xi(x)$ in the position representation.
From Eq.~\eqref{eq:wav} we immediately get
\begin{equation}
\xi(x)=\sum_{j=1}^{n}a_j\, \xi_1(x-(j-1)d),  
\end{equation}
where
\begin{equation}
\xi_1(x) = \frac{1}{\sqrt{2\pi}} \mathrm{PV}\!\! \int\frac{F(k)}{E-\omega(k)}\e^{\ii k x }\,\dd k.
\end{equation}
By assuming that $F(k)$ is real  and  has the same properties of $f(k)$ in (i)--(iv) of Appendix~\ref{app:self}, we can write
\begin{equation}
\xi_1(x) = \frac{1}{2} \big( \Xi(E+\ii 0) + \Xi(E-\ii 0) \big),
\label{eq:C3}
\end{equation}
where
\begin{equation}
\Xi(z,x) =  \frac{1}{\sqrt{2\pi}} \int\frac{F(k)}{z-\omega(k)}\e^{\ii k |x| }\,\dd k,
\end{equation}
can be computed by the residue theorem. The computation is a carbon copy of the computation of the self-energy in Appendix~\ref{app:self}, and by replacing $f(k)$ with $F(k)/\sqrt{2\pi}$ in \eqref{eq:B23}--\eqref{eq:B25} one has
\begin{equation}
\Xi(E+\ii 0) = - \ii \, W(E) \, \e^{\ii k(E) | x | } +  \eta(x) ,
\label{eq:C5}
\end{equation}
with a residue term
\begin{equation}
W(E)= \sqrt{2 \pi} \, \frac{F(k(E))}{\omega'(k(E))},
\end{equation}
and a (real valued) contour contribution
\begin{equation}\label{eq:defeta}
\eta(x) = \e^{- m  | x | }  \frac{1}{\sqrt{2\pi}} \int_{\mathbb{R}}\frac{F(k+\ii m)}{E- \omega(k+\ii m)} \, 
  \e^{\ii  k |x| }  \,\dd k .
\end{equation}
Analogously one gets
\begin{equation}
\Xi(E-\ii 0) =  \ii \, W(E) \, \e^{-\ii k(E) | x | } +  \eta(x) ,
\label{eq:C8}
\end{equation}
since for $z= E - \ii \delta$ the pole with positive imaginary part is $-\kappa(z)$ and $\omega'(-k(E))=-\omega'(k(E)$.

Finally, by plugging \eqref{eq:C5} and \eqref{eq:C8} into \eqref{eq:C3} we obtain
\begin{equation}
\xi_1(x) =  W(E)  \sin \bigl( k(E) | x |\bigr) + \eta(x),
\label{eq:C3bis}
\end{equation}
with $\eta(x)$ real and of order $O(\e^{- m  | x | } )$, that is  the expression~\eqref{eq:singleboson} in the main text.

\section{Calculation of the BICs}\label{app:calculation}
In this appendix we will evaluate explicitly the bound states in the continuum in our model, thus proving the results reported in Section~\ref{sec:evaluation}. The appendix is organized as follows:
\begin{itemize}
	\item in Appendix~\ref{subsec:strategy} we rephrase the problem in a convenient form;
	\item in Appendix~\ref{subsec:parity} we show that, because of the presence of corrections $\beta_{j-\ell}(E)$ to the propagator, BICs are predicted to have a definite parity;
	\item in Appendix~\ref{subsec:degen} we briefly revise the problem in the approximation in which all contributions $\beta_{j-\ell}(E)$, $j\neq\ell$, to the propagator in Eq.~\eqref{eq:propagator} are neglected: in this approximation, BICs emerge at \textit{resonant} energies $E=E_\nu$, $\nu=0,1,\dots$, and for each of these values there is an $(n-1)$-dimensional space of degenerate BICs that are ``switched on" for the same value of the excitation energy $\varepsilon$ of the emitters;
	\item in Appendix~\ref{subsec:nondegen} we refine our analysis by taking into account the largest correction $\beta_1(E)$ to the propagator, showing that this correction alone is sufficient to lift the degeneracy: while BICs still emerge (nearly) at resonant energies $E=E_\nu$, we will now be able to distinguish $n-1$ nondegenerate BICs, corresponding to distinct (albeit close) values of $\varepsilon$ and to well-defined collective structures.
\end{itemize}

\subsection{General strategy}\label{subsec:strategy}
Before getting started, let us rephrase Eqs.~\eqref{eq:BICs1}--\eqref{eq:BICs2} in a more compact way which will turn largely useful for our purposes. We start by defining the two quantities
\begin{equation}\label{eq:chi}
	b_j(E)=\frac{\beta_j(E)}{Z(E)},\qquad \chi(E)=\frac{\varepsilon-E}{Z(E)}+b_0(E),
\end{equation}
with $Z(E)$ as in Eq.~\eqref{eq:ze}. The propagator matrix $\mathrm{G}^{-1}(E)$ in~\eqref{eq:propagator} can be written as
\begin{equation}\label{eq:propa}
\mathrm{G}^{-1}(E)=- Z(E)\,\Bigl( A\big(k(E)d, b (E)\big)-\chi(E)\mathrm{I}_n\Bigr),
\end{equation}
where we introduce the matrix $\mathrm{A}(\theta,b)$ by
\begin{equation} \label{rearrProp}
\mathrm{A}_{j\ell}(\theta, b )=\begin{cases}
\ii &j=\ell\\
\ii \e^{\ii |j-\ell|\theta} - b_{|j-\ell |} &j\neq \ell
\end{cases},
\end{equation}
for $j,\ell=1,\dots,n$, $ b =(b_1,\dots,b_{n-1})$. It is immediate to show that, by Eq.~\eqref{eq:propa}, the two equations~\eqref{eq:BICs1}--\eqref{eq:BICs2} correspond to an $E$-dependent eigenvalue and eigenvector problem for the matrix $\mathrm{A}(\theta,b)$, namely
\begin{equation}
\begin{dcases}
\det\Bigl[A\Bigl(k(E)d,b(E)\Bigr)-\chi(E)\mathrm{I}_n\Bigr]=0;\\
A\Bigl(k(E)d,b(E)\Bigr)\bm{a}=\chi(E)\bm{a}.
\end{dcases}
\end{equation}
In this way, the evaluation of the BICs can be performed as follows:
\begin{enumerate}
	\item we search for the \textit{real} eigenvalues $\tilde{\chi}(\theta, b )$ of the matrix $\mathrm{A}(\theta, b )$,
	evaluate the corresponding eigenvectors $\bm{a} $ and check that the constraints~\eqref{eq-physicalcondition}, with $k(E)=\theta/d$, are satisfied;
	\item we obtain the energies $E$ of the corresponding BIC by solving the equation
	\begin{equation}
	\chi(E)=\tilde{\chi}\bigl(k(E)d, b (E)\bigr),
	\end{equation}
	with $\chi(E)$ as in Eq.~\eqref{eq:chi};
	\item finally, we obtain the photon amplitude of the eigenfunction by substituting the values of $E$ and $\bm{a}$ in~\eqref{eq:singleboson} .
\end{enumerate}

\subsection{Symmetry properties of the BICs}\label{subsec:parity}

Even without explicitly solving the aforementioned problem, a fundamental property of BICs in our system can be inferred by the structure of the matrix $\mathrm{A}(\theta, b )$: indeed, for all $j,\ell=1,\dots,n$,
\begin{equation}
\mathrm{A}_{j\ell}(\theta, b )=\mathrm{A}_{|j-\ell|}(\theta, b ),
\end{equation}
i.e., its entries are a function of the distance from the main diagonal. This is an immediate consequence of the fact that the Hamiltonian~\eqref{Hint} is parity-invariant and that the emitters are equally spaced. Consequently, $\mathrm{A}(\theta, b )$ is \textit{centrosymmetric}, i.e. satisfies the property
\begin{equation}
[A,J_n]=0,
\end{equation}
where $J_n$ is the \textit{exchange matrix}, i.e.\ the matrix having ones on the counterdiagonal
\begin{equation}
[J_n]_{j\ell}=\begin{cases}
1 &\text{if }\ell=n-j+1 \displaystyle \\
0 &\text{otherwise} \displaystyle
\end{cases}
\end{equation}
and which admits $\pm1$ as eigenvalues, with corresponding eigenvectors $(a_1, a_2, \dots, a_n)$ satisfying 
\begin{equation}
a_{n-j+1}=a_j \qquad \text{for } j=1,\dots,n,
\label{eq:sim}
\end{equation}
or
\begin{equation}
a_{n-j+1}=-a_j,\qquad  \text{for } j=1,\dots,n ,
\label{eq:antisim}
\end{equation}
respectively.
As a consequence, $J_n$ and $A$ always share a common basis of eigenvectors, which means that $A$ always admits a basis of eigenvectors that are either centrally symmetric or antisymmetric. Moreover, \textit{if} the eigenvalues of $A$ are nondegenerate, then \textit{necessarily} its corresponding eigenvectors will be centrally symmetric or antisymmetric. Physically, this is caused by the presence of tails of the photon wavefunctions, due to the fact that the dispersion relation is generally bounded from below and nonlinear. Though exponentially suppressed in the interatomic distance, these tails couple with all the emitters of the configuration, generally causing instability unless proper symmetry conditions are fulfilled \cite{bic}. 

Finally, we remark that the property discussed above is \textit{exact}. While the explicit calculation of the BICs in Appendix~\ref{subsec:nondegen} is performed under a nearest-neighbor approximation (and does, indeed, yield eigenstates with definite symmetry), taking into account higher-order terms, or even the full structure of the propagator, would not spoil this property, which is thus more fundamental.

\subsection{The degenerate case: resonance condition}\label{subsec:degen}

As discussed in the main text, the terms $b_{j-\ell}$ in the matrix~\eqref{rearrProp} are exponentially suppressed in $|j-\ell|md$, and are therefore small in the physically meaningful regime.
As a first approximation, one may simply discard them
and proceed with the study of the eigensystem of the matrix
\begin{equation}
\mathrm{A}_{j\ell}(\theta,0)=
\ii \,\e^{\ii |j-\ell|\theta}.
\end{equation}
This matrix admits real eigenvalues only if $\theta=\nu\pi$ for some $\nu\in\mathbb{N}$, when it becomes the rank-one matrix~\cite{bic}
\begin{equation}\label{eq:propres0}
\mathrm{A}(\nu\pi, 0)= \ii \,\bm{u}_\nu \bm{u}_\nu^\intercal,
\end{equation}
with 
\begin{equation}
\bm{u}_\nu =\bigl(1,(-1)^\nu,1,(-1)^\nu,\dots\bigr)^\intercal .
\end{equation}
The spectrum of $\mathrm{A}(\nu\pi,\bm{0})$ is thus composed of
\begin{itemize}
	\item the simple eigenvalue $\tilde{\chi}=\ii n$, associated with the one-dimensional eigenspace spanned by  $\bm{u}_\nu$;
	
	\item the $(n-1)$-degenerate eigenvalue $\tilde{\chi}=0$, whose eigenspace contains all the atomic amplitudes $\bm{a}$ orthogonal to $\bm{u}_\nu$:
	\begin{equation}
	\label{eq:physicalplus0}
	\bm{u}_\nu \cdot \bm{a}=0.
	\end{equation}	
\end{itemize}
While the eigenvector $\bm{u}_\nu$ corresponds to an unstable (superradiant) state of the system, the eigenvectors associated with the zero eigenvalue correspond to stable configurations of the system, which emerge if and only if
\begin{equation}
\begin{cases}
k(E)d=\nu\pi, \quad (\nu\in\mathbb{N})\\
\chi(E)=0.
\end{cases}
\end{equation}
The first equation is a \textit{resonance condition}: the emitter spacing $d$ must be an integer multiple of the half-wavelength $\pi/k(E)$ corresponding to the energy $E$. Hence, if we define the \emph{resonant energies}
\begin{equation}
E_\nu=\omega\left(\frac{\nu\pi}{d}\right), \quad (\nu\in\mathbb{N})
\label{eq:resonance}
\end{equation}
a BIC with energy $E$ exists if and only if
\begin{equation}
\begin{cases}
E=E_\nu,\\
\chi(E_\nu)=0 .
\end{cases}
\end{equation}
Observe that by Eq.~\eqref{eq:chi} the second condition reads
\begin{equation}
\varepsilon=E_\nu -  Z(E_\nu)b_0(E_\nu).
\end{equation}
As a consequence, a BIC emerges in the spectrum whenever its energy $E_\nu$ equals
the excitation energy $\varepsilon$ of the emitters (see Fig.\ \ref{fig:systemn}) plus a coupling-dependent correction, that is small in the perturbative regime.

We remark that  condition~\eqref{eq:physicalplus0} reads, explicitly,
\begin{equation}\label{eq:physicalplus}
\sum_{j=1}^n (-1)^{j\nu} a_j=0.
\end{equation}
Due to degeneracy, no symmetry condition is imposed \textit{a priori} onto the eigenvectors. Therefore, states in which the excitation is shared only by two consecutive emitters with opposite amplitudes, such as those depicted in Fig.~\ref{fig:tails}, represent eigenstates as valid as the ones with central symmetry, provided the evanescent fields are neglected.

\subsection{Nearest-neighbor approximation: degeneracy lifting and excitation waves} \label{subsec:nondegen}
As discussed in Subsection~\ref{subsec:nearest}, the degenerate situation outlined in Appendix~\ref{subsec:degen} is drastically modified when one takes into account the full structure of the self-energy and the propagator, and it suffices to only take into account the correction $b_1(E)$ in order to understand the ``direction'' in which the degeneracy is broken. 

In this case, an interesting structure emerges: adding the nearest-neighbor approximation to Eq.~\eqref{eq:propres0}, we obtain
\begin{equation}\label{eq:propres}
A (\nu\pi,b)= \ii \bm{u}_\nu  \bm{u}_\nu^\intercal - b_1\Delta_n ,
\end{equation}
where $\Delta_n$ is the $n\times n$ matrix in Eq.~\eqref{eq:laplace}, whose properties are studied in Appendix~\ref{app:laplace}. Eq.~\eqref{eq:propres} means that, when taking into account the role of $b_1$, the eigenproblem studied in Appendix~\ref{subsec:degen} is perturbed by a term which is proportional to the adjacency matrix $\Delta_n$. Since the unperturbed matrix $\mathrm{A}(\nu\pi,0)$ is rank-one, and thus highly degenerate, its eigensystem will be crucially affected by the structure of the perturbation, even for small $b_1$.

Here, we are interested in the spectrum and eigenvectors of the matrix~\eqref{eq:propres} in the limit $b_1\to0$, since the nearest-neighbor approximation is physically valid when $b_1$ is small. It is worth noticing that the obtained result have an interesting universal character, being independent of the specific form of $b_1$.

\begin{figure*}
	\centering
	\begin{subfigure}[t]{0.5\linewidth}
		\includegraphics[scale=0.43]{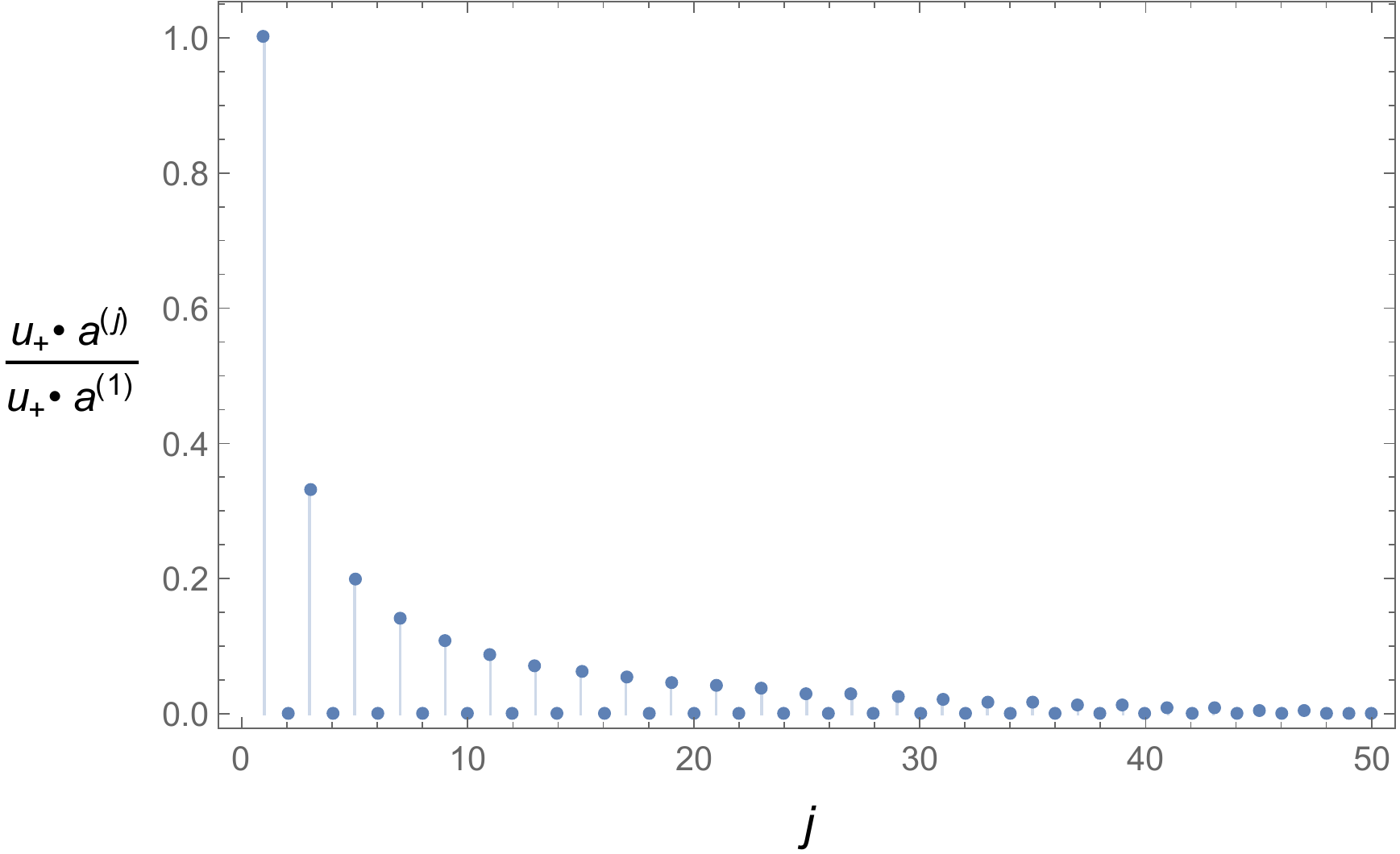}
		\caption{$n=50$, even $\nu$}
	\end{subfigure}
	\begin{subfigure}[t]{0.5\linewidth}
		\includegraphics[scale=0.43]{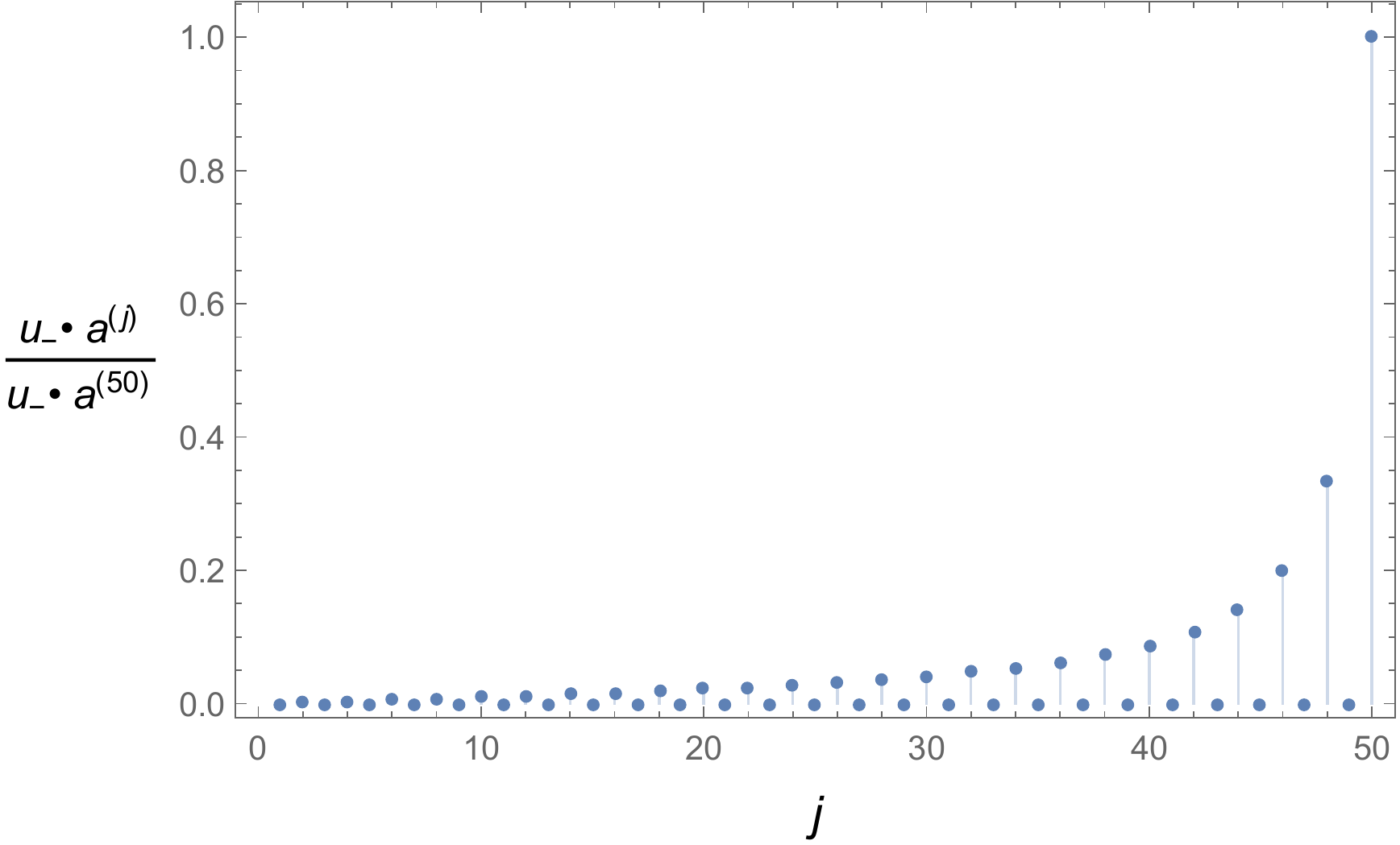}
		\caption{ $n=50$, odd $\nu$}
	\end{subfigure}\vspace{0.3cm}
	\begin{subfigure}[t]{0.5\linewidth}
		\includegraphics[scale=0.43]{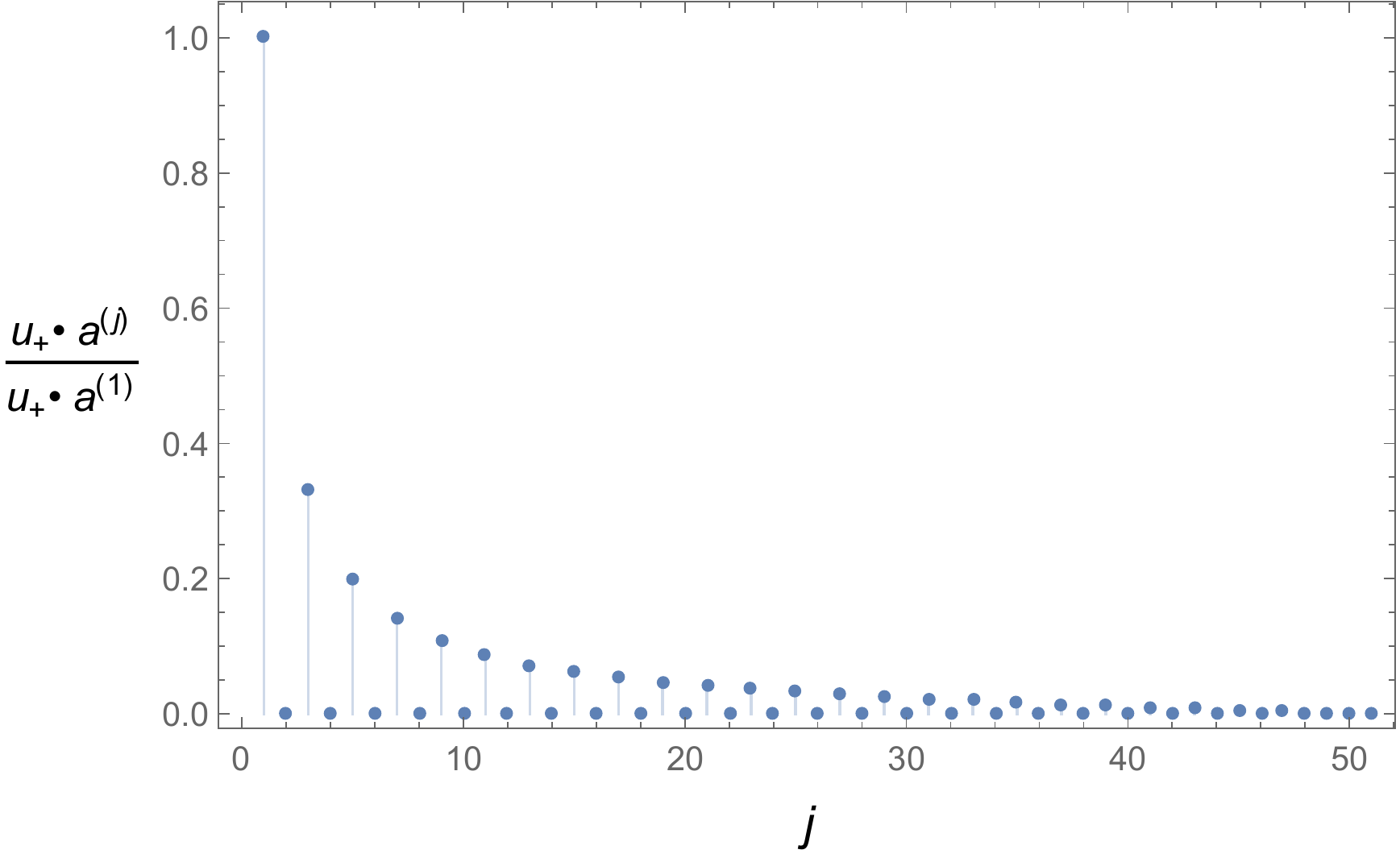}
		\caption{ $n=51$, even $\nu$}
	\end{subfigure}
	\begin{subfigure}[t]{0.5\linewidth}
		\includegraphics[scale=0.43]{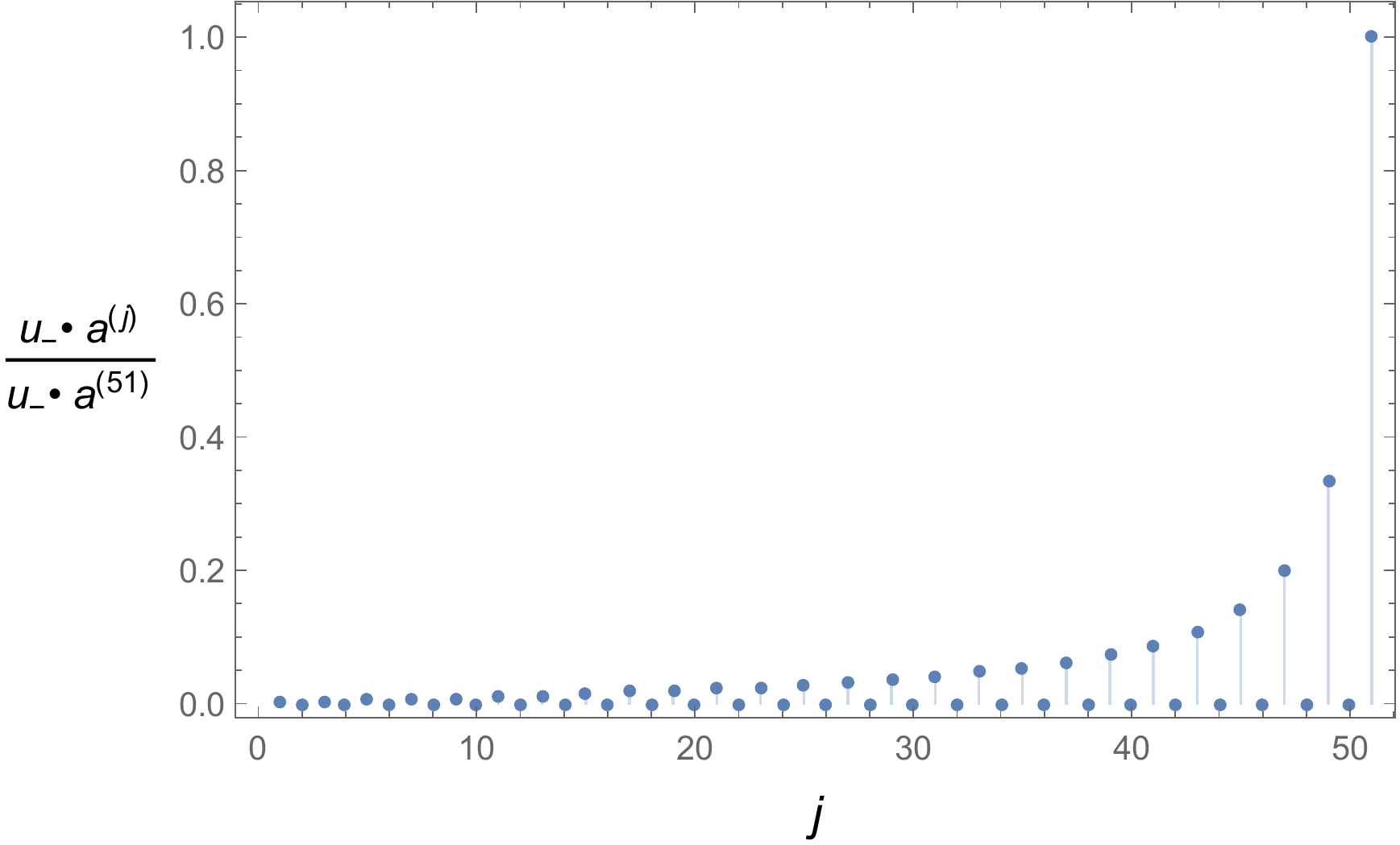}
		\caption{ $n=51$, odd $\nu$}
	\end{subfigure}\vspace{0.3cm}
	\captionsetup{justification=justified, singlelinecheck=false}
	\caption{Scalar products $\bm{u}_\nu\cdot\bm{a}^{(j)}$ for $n=50$ and $n=51$ vs $j$. All values are normalized with respect to the maximal value of the product. Nonvanishing values converge to zero either for small or large $j$: smaller values of the product correspond to smaller deformations with respect to the corresponding sinusoidal shape. }
	\label{fig:resonance}
\end{figure*}

\begin{figure*}[h!t!]\centering
	\begin{subfigure}{0.52\linewidth}
		\includegraphics[scale=0.55]{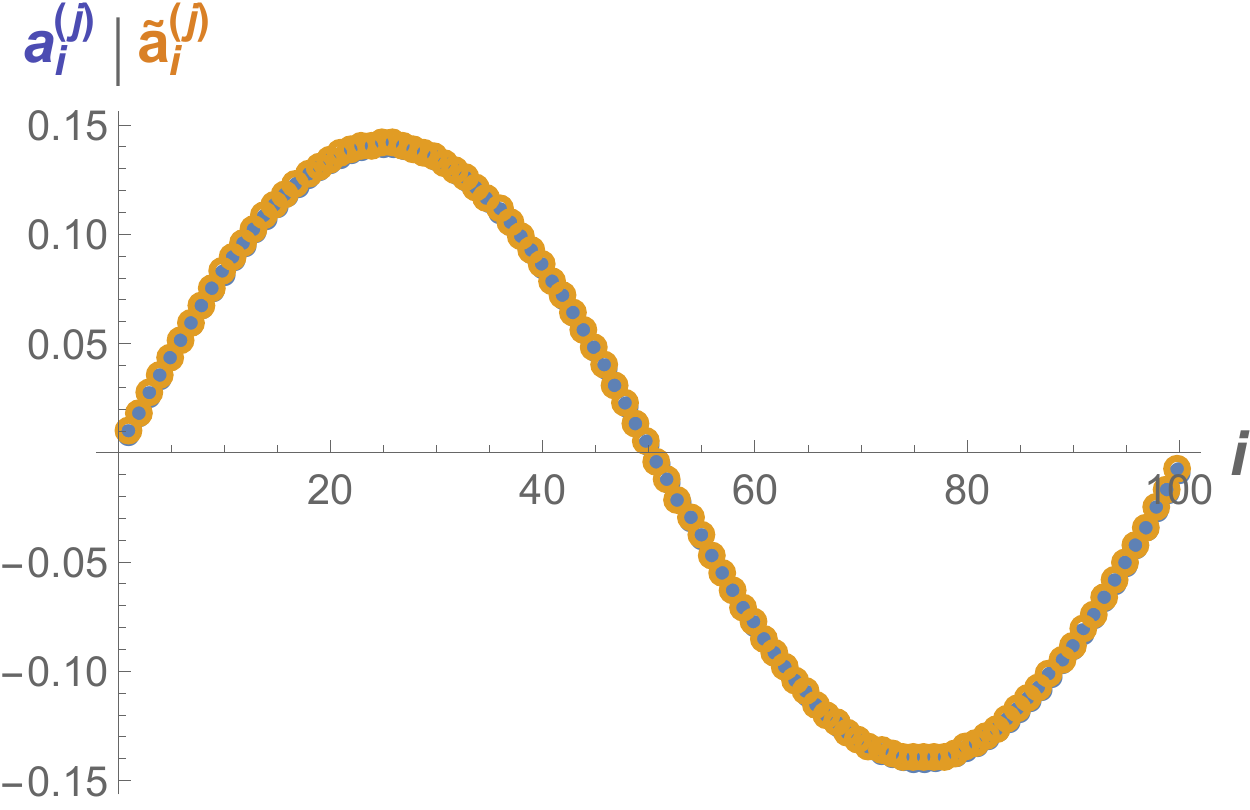}
		\caption{ $j=2$, exact excitation wave}
	\end{subfigure}%
	\begin{subfigure}{0.52\linewidth}
		\includegraphics[scale=0.55]{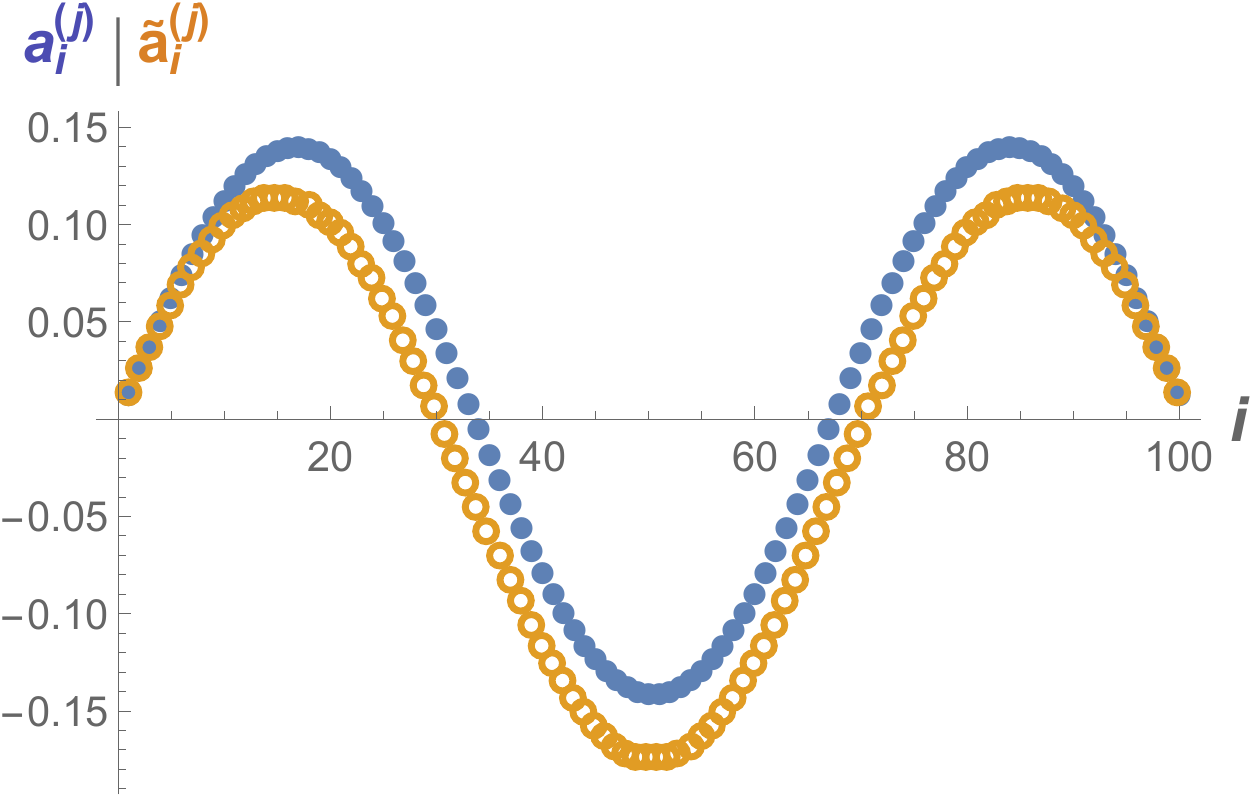}
		\caption{ $j=3$, deformed excitation wave}
	\end{subfigure}\vspace{0.3cm}
	\begin{subfigure}{0.52\linewidth}
		\includegraphics[scale=0.55]{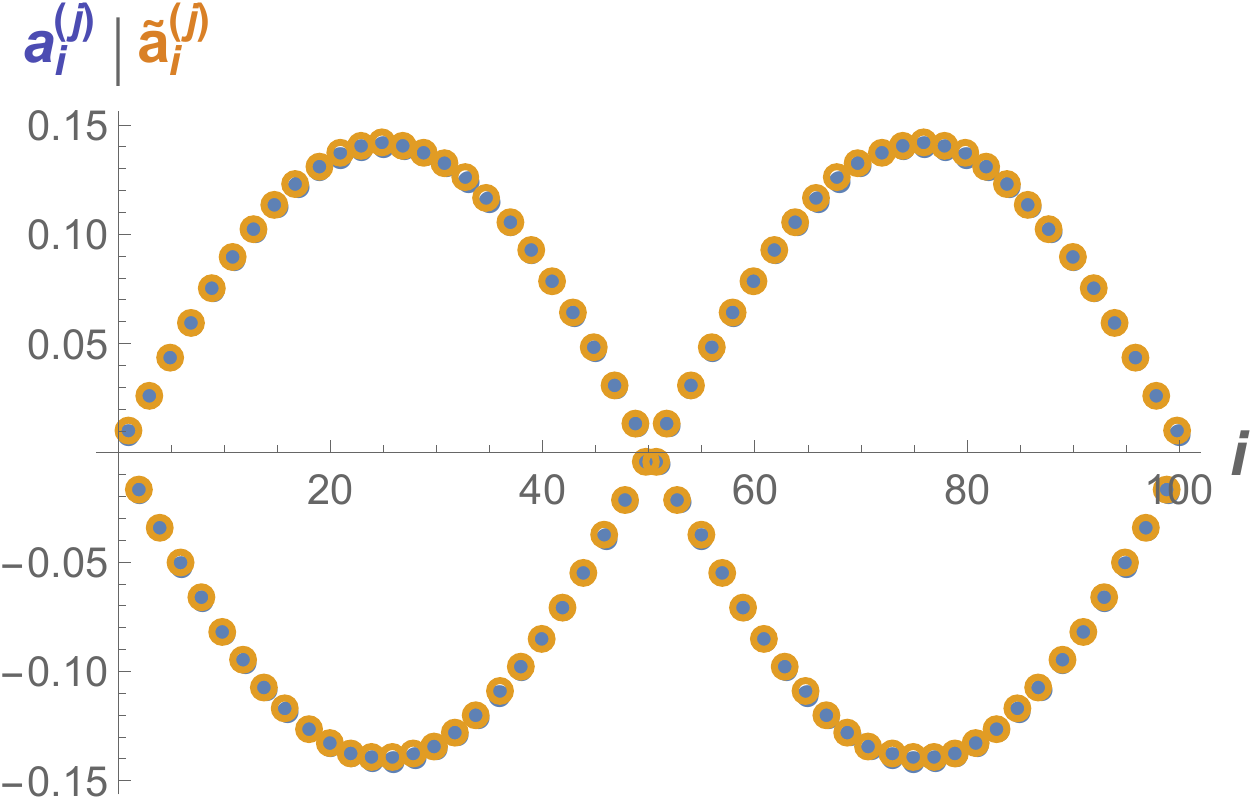}
		\caption{ $j=99$, deformed excitation wave}
	\end{subfigure}%
	\begin{subfigure}{0.52\linewidth}
		\includegraphics[scale=0.55]{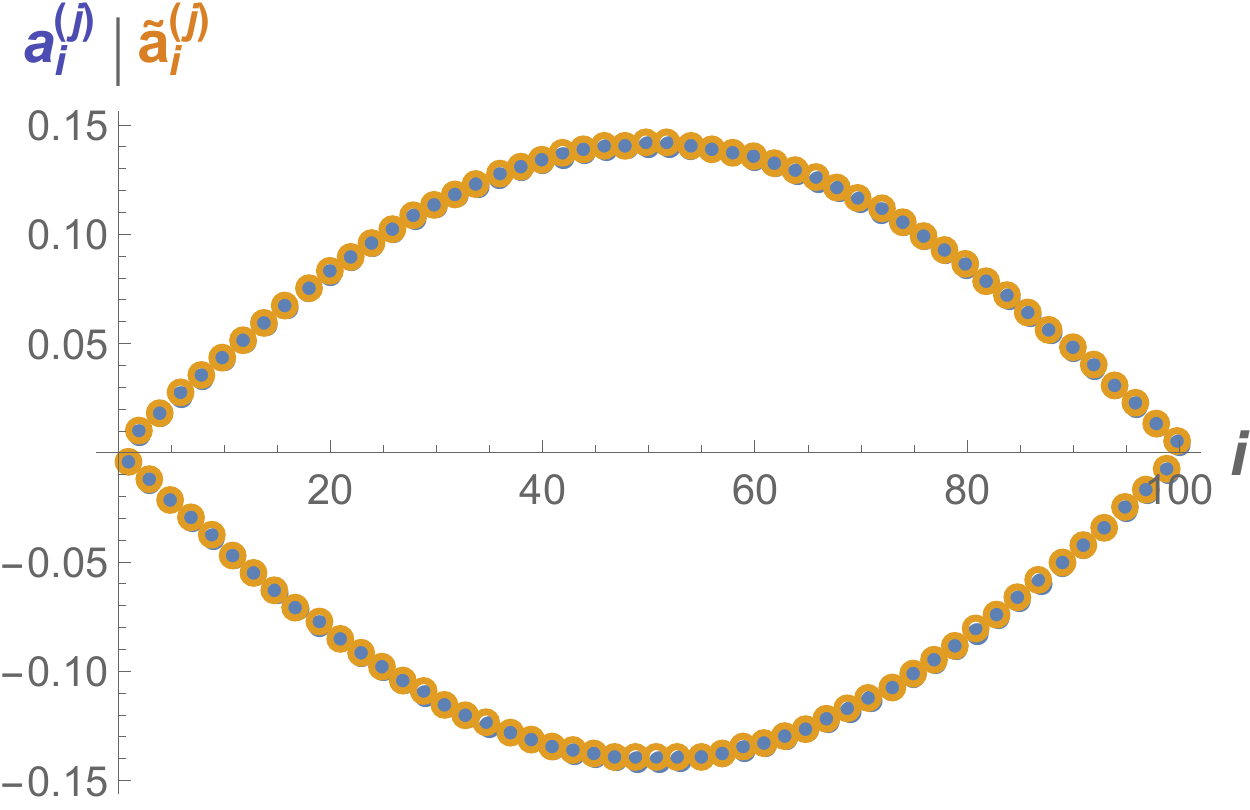}
		\caption{ $j=100$, exact excitation wave}
	\end{subfigure}
	\vspace{0.3cm}
	\captionsetup{justification=justified, singlelinecheck=false}
	\caption{The numerically determined eigenstates $\tilde{\bm{a}}_\nu^{( j)}$ (open orange circles) of the matrix $\mathrm{A}(\nu\pi,b)$, for $n=100$ emitters and even $\nu$, are compared with the excitation waves $\bm{a}^{(j)}$ (full blue circles), eigenstates of the nearest-neighbor adjacency matrix $\Delta_n$. The upper and lower panels represent acoustic and optical waves, respectively. Eigenstates with even $j$ coincide with the excitation waves, while those with odd $j$ are deformed. The effect of deformation is more relevant for acoustic waves than for optical waves. Notice that the excitation wave with $j=1$ corresponds to an eigenvalue with large imaginary part. The amplitude vectors $\bm{a}^{(j)}$ and $\tilde{\bm{a}}_\nu^{( j)}$ are normalized to their respective square norms.}
	\label{fig:deform}
\end{figure*}

\begin{figure*}
	\centering
	\begin{subfigure}{0.52\linewidth}
		\includegraphics[scale=0.55]{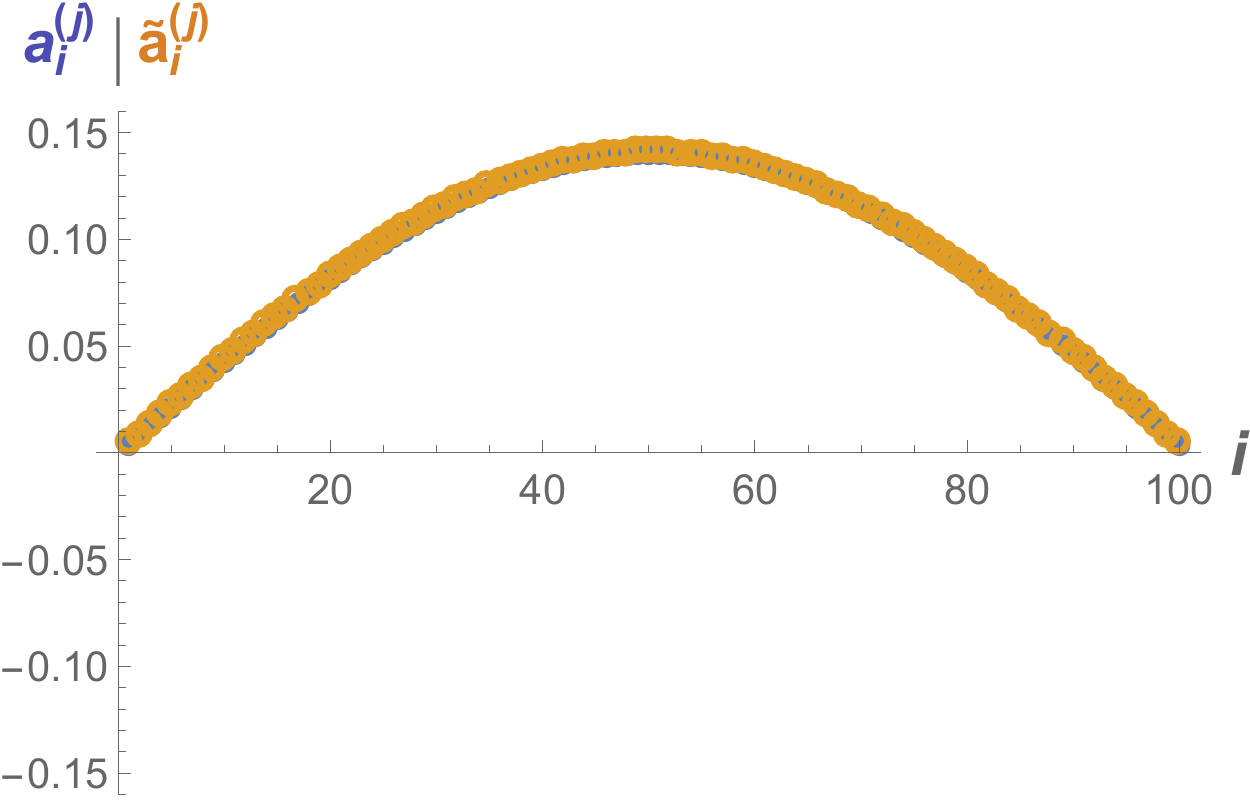}
		\caption{ $j=1$, exact excitation wave}
	\end{subfigure}%
	\begin{subfigure}{0.52\linewidth}
		\includegraphics[scale=0.55]{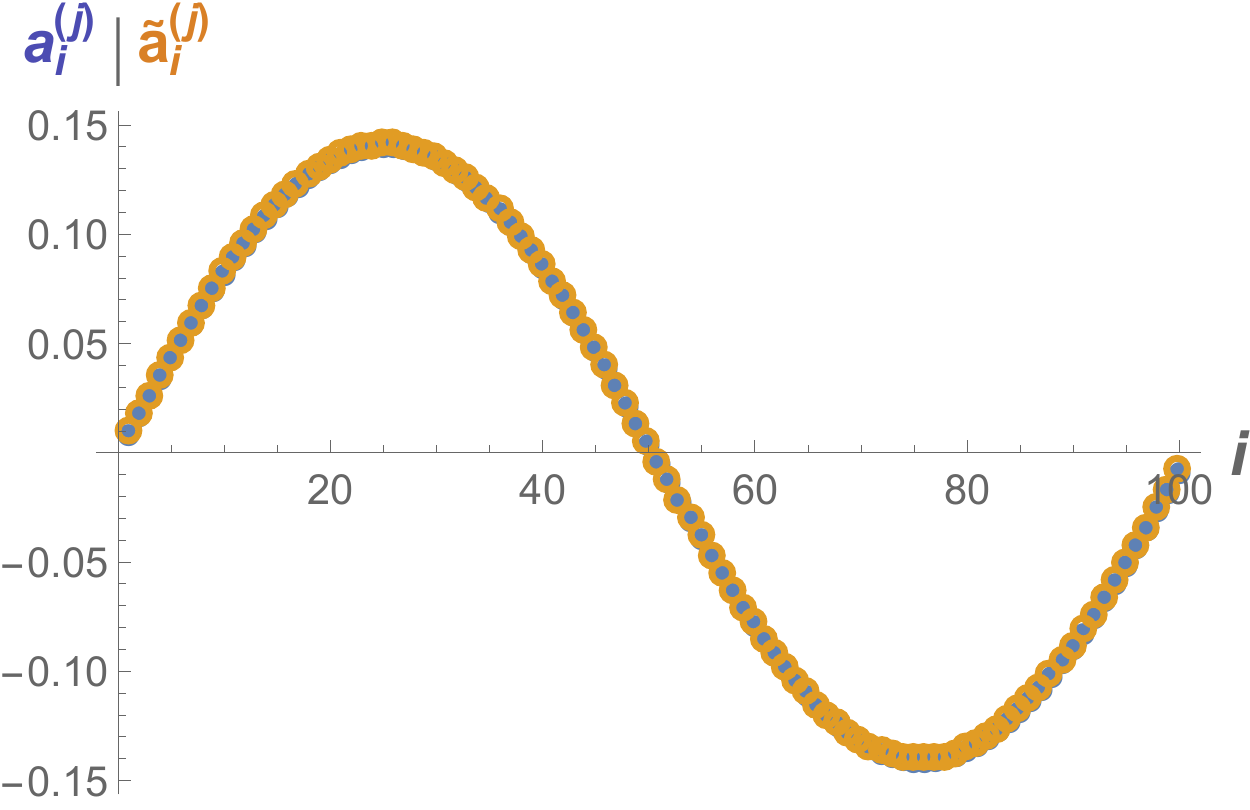}
		\caption{ $j=2$, deformed excitation wave}
	\end{subfigure}\vspace{0.3cm}
	\begin{subfigure}{0.52\linewidth}
		\includegraphics[scale=0.55]{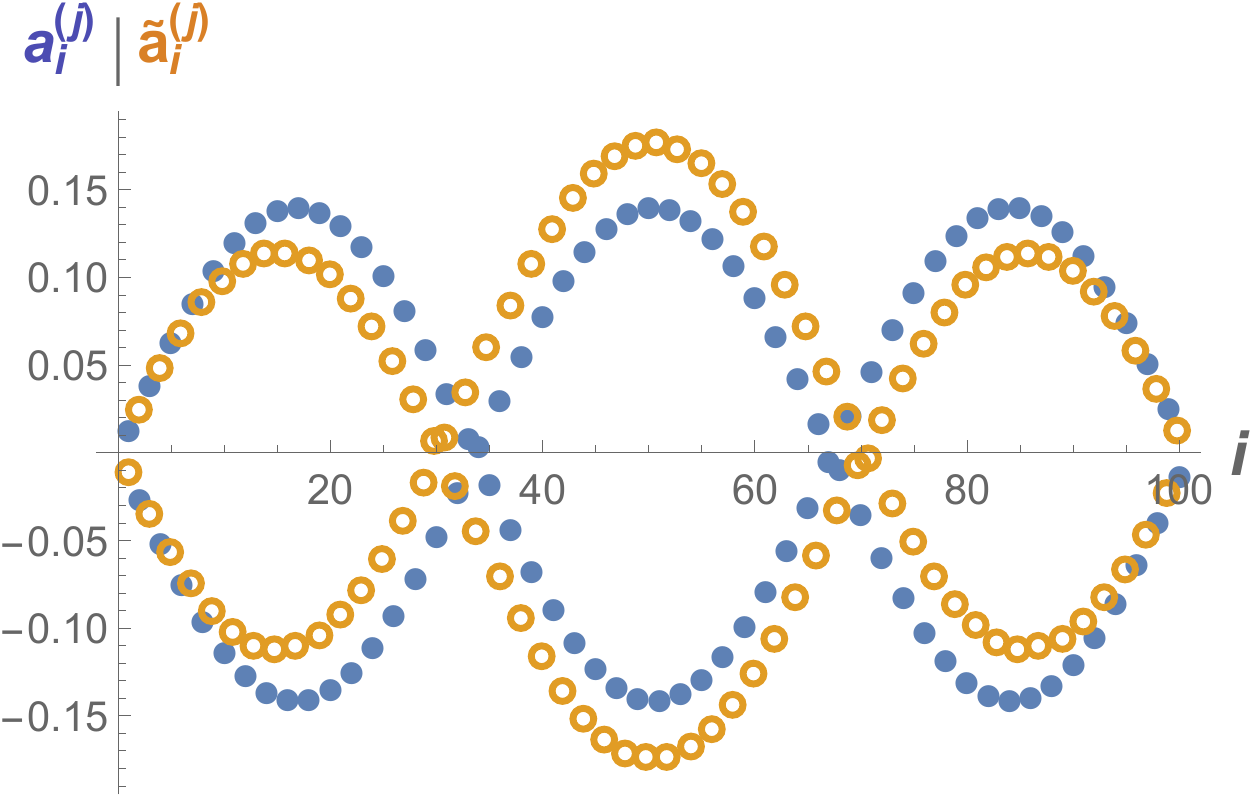}
		\caption{ $j=98$, deformed excitation wave}
	\end{subfigure}%
	\begin{subfigure}{0.52\linewidth}
		\includegraphics[scale=0.55]{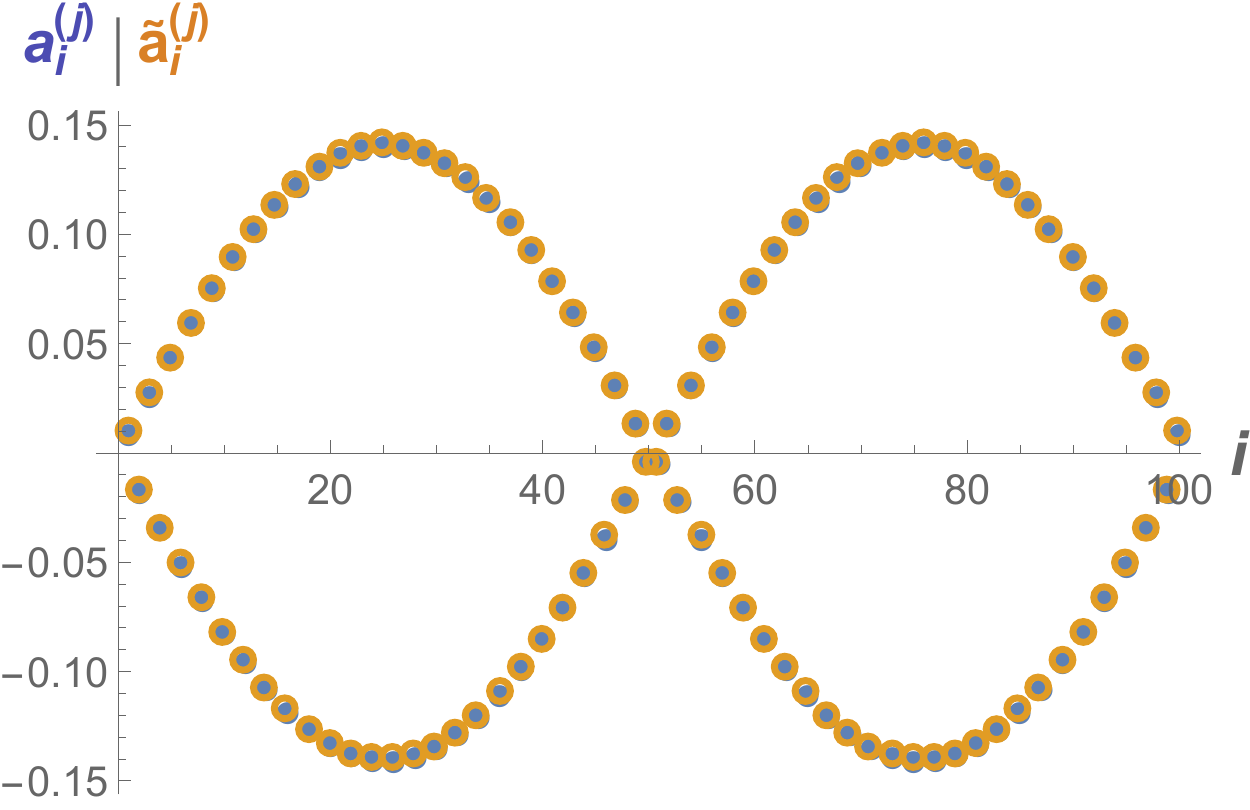}
		\caption{ $j=99$, exact excitation wave}
	\end{subfigure}\vspace{0.3cm}
	\captionsetup{justification=justified, singlelinecheck=false}
	\caption{The numerically determined eigenstates $\tilde{\bm{a}}_\nu^{( j)}$ (open orange circles) of the matrix $\mathrm{A}(\nu\pi,b)$, for $n=100$ emitters and odd $\nu$, are compared with the excitation waves $\bm{a}^{(j)}$ (full blue circles), eigenstates of the nearest-neighbor adjacency matrix $\Delta_n$. The upper and lower panels represent acoustic and optical waves, respectively. Eigenstates with odd $j$ coincide with the excitation waves, while those with even $j$ are deformed. The effect of deformation is more relevant for optical waves than for acoustic waves. Notice that the excitation wave with $j=100$ corresponds to an eigenvalue with large imaginary part. The amplitude vectors $\bm{a}^{(j)}$ and $\tilde{\bm{a}}_\nu^{( j)}$ are normalized to their respective square norms.}
	\label{fig:deform2}
\end{figure*}
As shown in Appendix~\ref{app:laplace}, the eigenvalues $\chi^{(j)}$ of the matrix $\Delta_n$, with $1\leq j\leq n$, can be evaluated exactly, see Eq.~\eqref{eq:eigvl} in the main text: these are the sinusoidal excitation waves.

By applying the propagator at resonance, $\theta=\nu\pi$, we get, 
\begin{equation}
\mathrm{A}(\nu\pi,b)\bm{a}^{(j)}= - b_1\chi^{(j)}\bm{a}^{(j)}+ \ii \, \left(\bm{u}_\nu \cdot \bm{a}^{(j)}\right) \bm{u}_\nu.
\end{equation}
The excitation waves $\bm{a}^{(j)}$ and the vector $\bm{u}_\nu$ are related as follows: 
\begin{itemize}
	\item if $n$ is even, then
	\begin{align}
	\bm{u}_\nu\cdot\bm{a}^{(j)} = 0 & \quad \text{for even } j+\nu \\ 
	\bm{u}_\nu\cdot\bm{a}^{(j)} \neq 0 & \quad \text{for odd } j+\nu 
	\end{align}
	\item if $n$ is odd, then
	\begin{align}
	\bm{u}_\nu\cdot\bm{a}^{(j)} = 0 & \quad \text{for even } j  \\ 
	\bm{u}_\nu\cdot\bm{a}^{(j)} \neq 0 & \quad \text{for odd } j
	\end{align}
\end{itemize}
Moreover, the products $ \bm{u}_\nu \cdot\bm{a}^{(j)}$, even when nonvanishing, tend to become negligible for large (small)  $j$, when $\nu$ is even (odd), as reported in Fig.~\ref{fig:resonance}. This is due to the fact neighboring excitation amplitudes in $\bm{a}^{(j)}$ tend to have the same sign for small $j$ and opposite sign for large $j$. Therefore, the scalar product with $\bm{u}_{\nu}$ for even $\nu$, which is a constant vector, tends to vanish for states with large $j$, while the product with $\bm{u}_{\nu}$ for odd $\nu$, which is a staggered vector of $\pm 1$, is minimal for small $j$.

As a consequence, 
\begin{itemize}
	\item if $n$ is even, then $\mathrm{A}(\nu\pi,b)$ admits as eigenvectors the excitation waves with even $j$ (respectively odd $j$) iff $\nu$ is even (respectively odd);
	\item if $n$ is odd, then $\mathrm{A}(\nu\pi,b)$ admits as eigenvectors the excitation waves with even $j$, for both even and odd $\nu$;
\end{itemize}
in the cases listed above, the excitation waves are also eigenvalues of the  at the resonance energy~$E_\nu$ given by \eqref{eq:resonance}.

Gathering all cases together, we observe, for every value of $n$ and $\nu$, that $\lfloor\frac{n}{2}\rfloor$ out of the $n$ excitation waves are also eigenvectors of the propagator matrix, and thus correspond, for some excitation energy $\varepsilon$ of the emitters, to a bound state in the continuum with resonant energy $E=E_\nu$. Specifically, an admissible BIC according to the aforementioned criterion is present in the spectrum of the Hamiltonian provided
\begin{equation}
\chi(E_\nu)=\chi^{(j)},
\end{equation}
and thus if and only if the excitation energy $\varepsilon$ of the emitters satisfies
\begin{equation}\label{eq:varepsilon}
\varepsilon=E_\nu - Z(E_\nu)\left(b_0(E_\nu)+b_1(E_\nu)\chi^{(j)}\right).
\end{equation}
Notice that, since the $\chi^{(j)}$'s are all distinct, in principle only one among the possible BICs will actually be a stable state for the array, even though, in practice, all such states are expected to be long-lived, with an $O(\e^{-md})$ lifetime. 

The remaining $\lceil\frac{n}{2}\rceil$ excitation waves are \textit{not} eigenstates of the propagator, since they do not satisfy the condition $\bm{u}_\nu\cdot\bm{a}^{(j)}=0$. 
A numerical analysis of the spectrum of the matrix $\mathrm{A}(\nu\pi,b)$ for small $b_1$ shows, besides the already discussed $\lfloor\frac{n}{2}\rfloor$ exact excitation waves, the following features:
\begin{itemize}
	\item One of the eigenvectors of $\mathrm{A}(\nu\pi,b)$ correspond to a complex eigenvalue with a large imaginary part. This state, characterized by $j=1$ for even $\nu$, and $j=n$ for odd $\nu$, represents a perturbation of the unstable eigenvector $\bm{u}_\nu$ of the matrix $\mathrm{A}(\nu\pi,0)$, with eigenvalue $\tilde{\chi}=\ii n$.
	\item The remaining $\lceil\frac{n}{2}\rceil-1$ eigenvectors are \textit{deformed} versions of the excitation waves that do not exactly satisfy the resonance condition. Their associated eigenvalues $\tilde{\chi}^{\pm,(j)}$ have real parts independent of $b_1$ and close to the eigenvalues $\chi^{(j)}$ in Eq.~\eqref{eq:eigvl}, while they also acquire an imaginary part (namely, a finite decay rate) that vanishes with $b_1$.
\end{itemize}
Summarizing, as reported in Table~\ref{tab:summ} in the main text and Figs.~\ref{fig:deform}-\ref{fig:deform2}, the $(n-1)$-dimensional degenerate eigenspace associated with the zero eigenvalue of the approximate propagator $\mathrm{A}(\nu\pi,0)$ breaks into $n-1$ nondegenerate eigenvectors, $\lfloor\frac{n}{2}\rfloor$ of which are the exact sinusoidal waves that are orthogonal to $\bm{u}_\nu $ (i.e., they satisfy the resonance condition at $\theta=\nu\pi$), while the remaining ones are ``deformed'' versions of the sinusoidal waves that have a small $[O(b_1)]$ but finite projection onto $\bm{u}_\nu$. The $j$th deformed excitation wave will be written as
\begin{equation}
\tilde{\bm{a}}_\nu^{( j)}=\bm{a}^{(j)}+\bm{\delta}_\nu^{( j)},
\end{equation}
with $\bm{\delta}_\nu^{( j)}$ being the deformation with respect to the $j$th exact sinusoidal wave. 

We also remark that, due to the behavior of the product $\bm{u}_\nu\cdot\bm{a}^{(j)}$, the size of the deformation of the wave crucially depends on $j$ in an opposite way for the two cases (see Fig.~\ref{fig:deform}):
\begin{itemize}
	\item for even $\nu$ the deformation $\bm{\delta}_\nu^{( j)}$ is larger for small $j$ (i.e. for acoustic waves), while it becomes negligible for large $j$ (i.e.\ for optical waves);
	\item conversely, for odd $\nu$, the deformation $\bm{\delta}_\nu^{( j)}$ is larger for large $j$ (i.e. for optical waves), while it becomes negligible for small $j$ (i.e.\ for acoustic waves).
\end{itemize}
This is a natural consequence of the fact that, as shown in Fig.~\ref{fig:resonance}, for off-resonant excitation waves, the term $ \bm{u}_\nu\cdot\bm{a}^{(j)}$ monotonically decays (increases) with $j$ for even (odd) $\nu$. 
In other words, the deformation will be relevant only for acoustic waves if $\nu$ is even and for optical waves if $\nu$ is odd, for every value of $n$.

As already mentioned, the deformed waves described above correspond to eigenvalues with a finite imaginary part, and therefore do not represent stable states. However, it is possible to find an actual BIC with energy $E\simeq E_{\nu}$ such that
\begin{equation}
k(E)d =\nu\pi+\delta k(E)\:d, \quad \text{ with }\delta\theta = O(b_1),
\end{equation}
close to each deformed wave. Moreover, since the imaginary parts of the deformed wave amplitudes are negligible, as $n$ increases they tend to coincide with very good approximation with the actual bound states, and the condition \eqref{eq:varepsilon} at which they appear in the spectrum is still valid. 

\section{Eigenvalues and eigenvectors of the adjacency matrix}\label{app:laplace}
Given $n\in\mathbb{N}$, let $\Delta_n$ be the adjacency matrix~\eqref{eq:laplace}. In this appendix we will
discuss the interpretation of the eigenvalue problem for the adjacency matrix $\Delta_n$ as  a discrete boundary value problem for a one-dimensional regular array, and evaluate its eigenvalues and eigenvectors.

We intend to solve the eigenvalue equation
\begin{equation}
\Delta_n \bm{a}  = \chi \bm{a}, 
\end{equation}
where  $\chi\in\mathbb{R}$ and $\bm{a} =(a_1,\dots,a_n)^\intercal\in\mathbb{C}^n$. For this purpose, let us consider an array of $n+2$ elements by adding two fictitious nodes $a_0$, $a_{n+1}$ at the extrema of the array $\bm{a} $. By construction, the eigenvalue problem for the matrix $\Delta_n$ is equivalent to a recurrence relation for the array $a_0,a_1,\dots,a_h,a_{h+1}$ with two boundary conditions at the extrema of the array:
\begin{equation}
	\begin{cases}
a_{\ell+1}=\chi\,a_\ell - a_{\ell-1}, \qquad\ell=1,\dots,n,\\
a_0= a_{n+1}=0,
	\end{cases}
	\label{eq:B2}
\end{equation}
which can be interpreted as a discrete version of the Helmholtz equation with vanishing (Dirichlet) boundary conditions.

The straightforward trigonometric identity
\begin{equation}
 \sin(\ell +1)\theta = 2 \cos \theta \, \sin \ell \theta - \sin(\ell -1) \theta  ,
\end{equation}
implies that if $\chi = 2 \cos \theta$, then $a_\ell =  \sin \ell \theta$ is a solution of the recurrence relations in~\eqref{eq:B2} for $\ell= 1, \dots, n$.
The boundary condition  at $\ell=0$, $a_0= 0$, is automatically satisfied, while the boundary condition at $\ell = n+1$,  $a_{n+1}= 0$ fixes the admissible values of $\theta$, and thus the eigenvalues of the original problem, namely
\begin{equation}
\sin (n+1)\theta = 0,
\end{equation}
that is $\theta^{(j)} = j \pi/(n+1)$ with $j=1,\dots n$.

Since  the adjacency $\Delta_n$ is a symmetric matrix, with $n$ real eigenvalues, we have obtained in this way its full spectrum: 
the eigenvalues $\chi^{(j)}$ and the eigenvectors $\bm{a}^{(j)}=(a^{(j)}_1,\dots,a^{(j)}_n)^\intercal$, $j=1,\dots,n$ of $\Delta_n$ read
\begin{align}
\chi^{(j)}=& \, 2\cos\left(\frac{j \pi }{n+1}\right),\\
a_\ell^{(j)}=& \frac{1}{\mathcal{N}_j}\sin\left(\ell \frac{j\pi}{n+1}\right), \label{eigODD}
\end{align}
where $\ell=1,\dots,n$, and $\mathcal{N}_j$ is a suitable normalization coefficient.

\begin{figure*}
	\centering
	\begin{subfigure}[t]{0.52\linewidth}
		\includegraphics[scale=0.43]{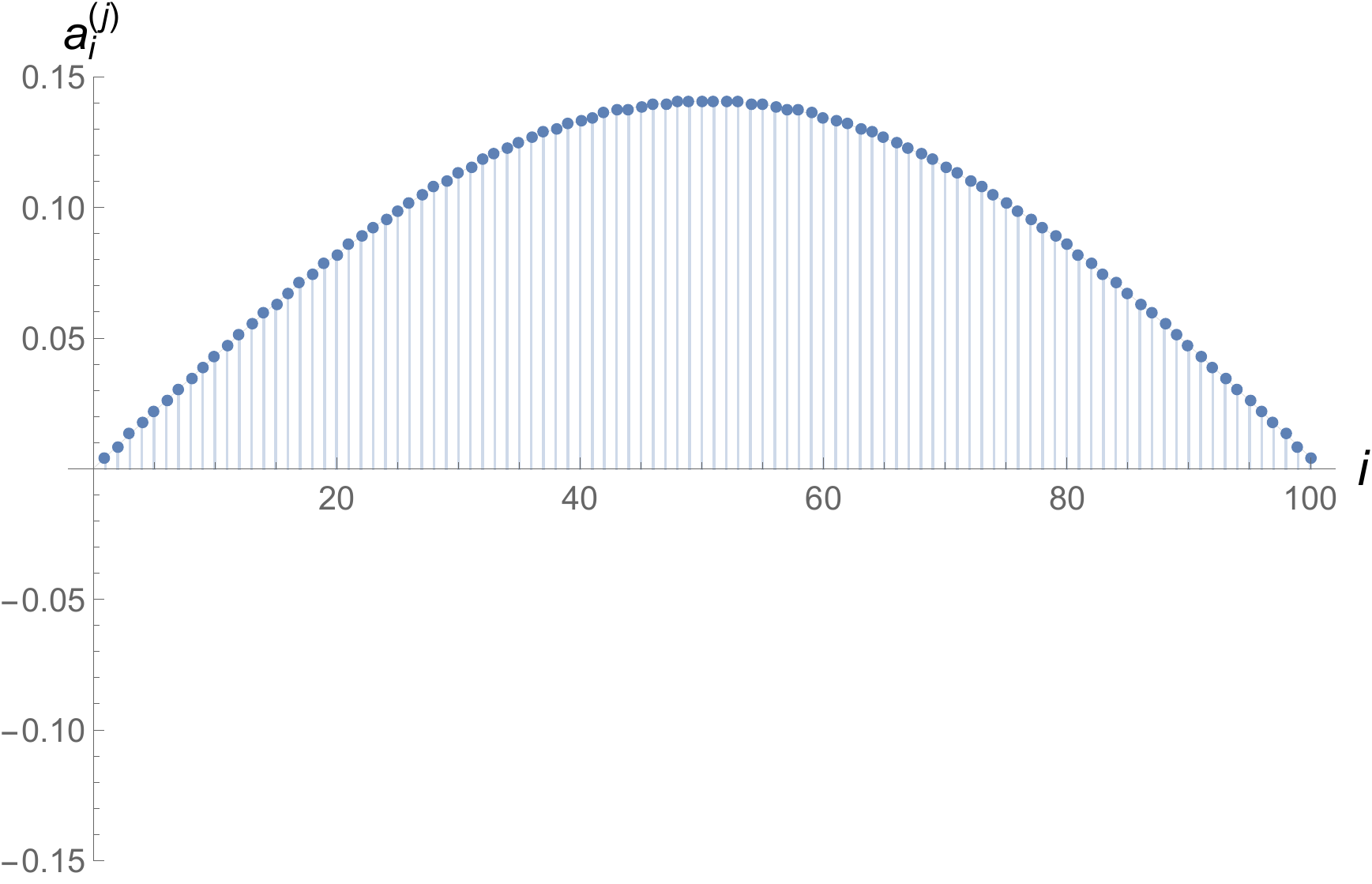}
		\caption{ $n=100$, $j=1$}
	\end{subfigure}
	\begin{subfigure}[t]{0.52\linewidth}
		\includegraphics[scale=0.43]{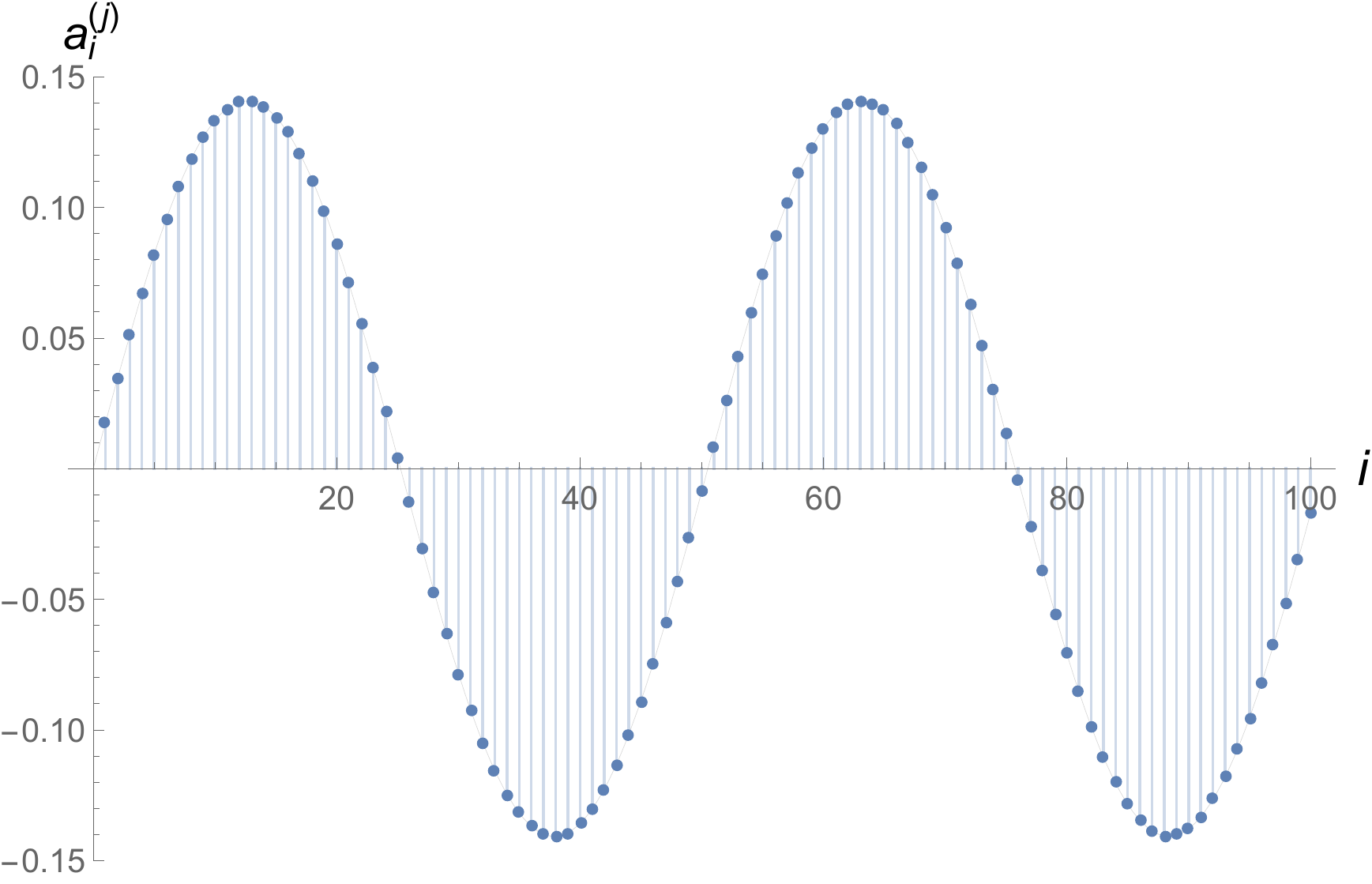}
		\caption{ $n=100$, $j=4$}
	\end{subfigure}\vspace{0.3cm}
	\begin{subfigure}[t]{0.52\linewidth}
		\includegraphics[scale=0.43]{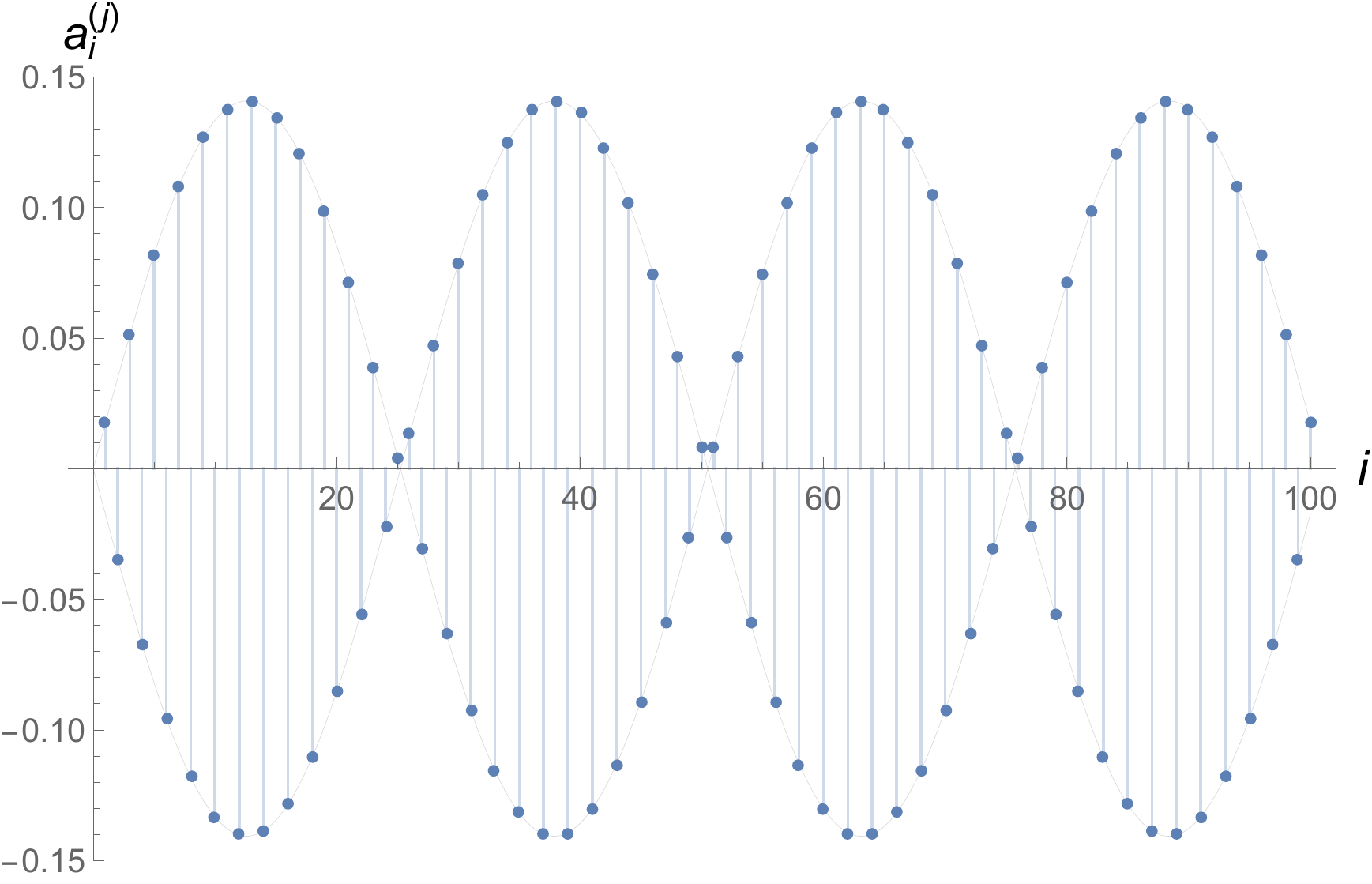}
		\caption{ $n=100$, $j=97$}
	\end{subfigure}
	\begin{subfigure}[t]{0.52\linewidth}
		\includegraphics[scale=0.43]{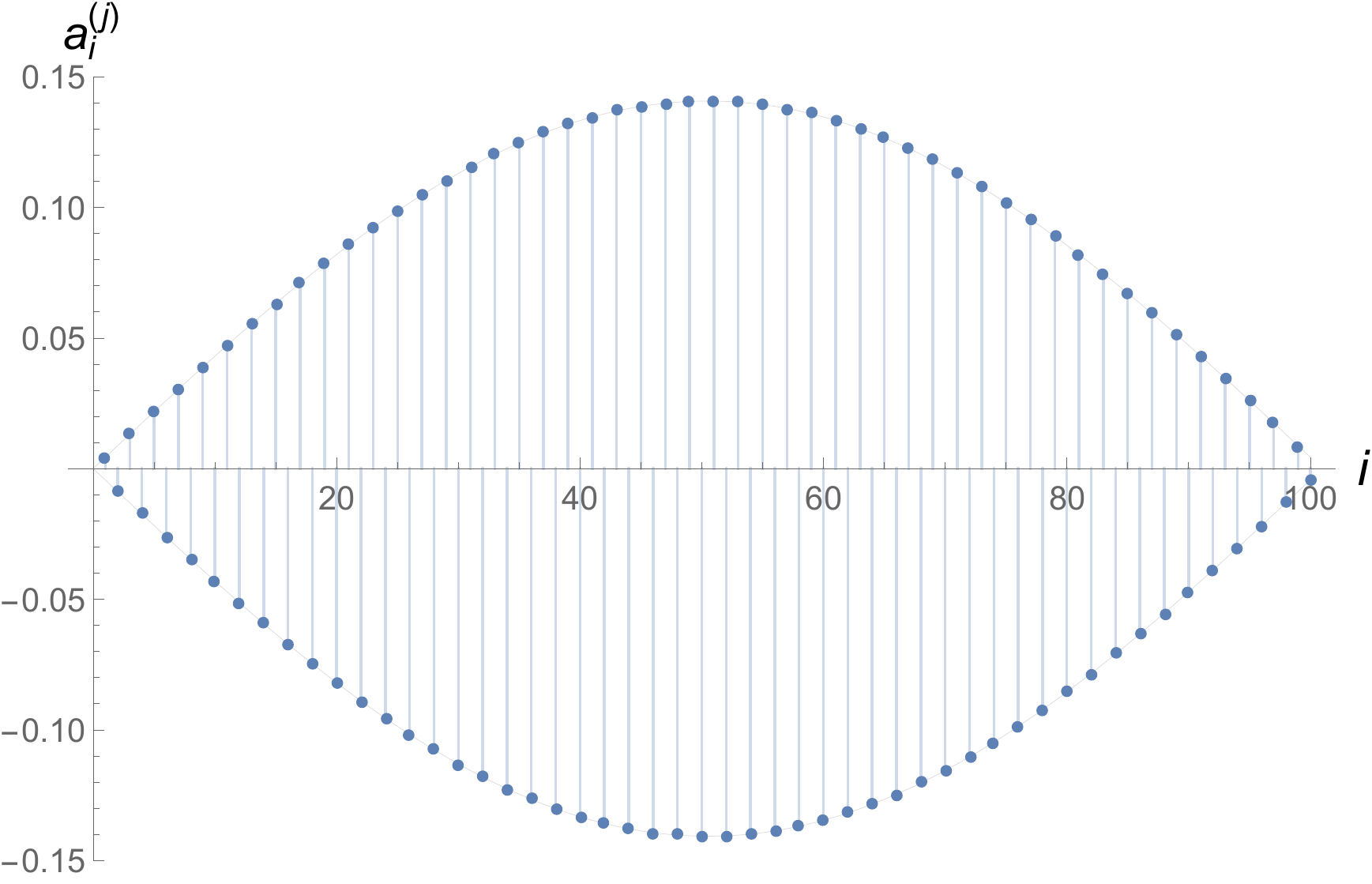}
		\caption{ $n=100$, $j=100$}
	\end{subfigure}
	\captionsetup{justification=justified, singlelinecheck=false}
	\caption{Excitation waves $\bm{a}^{(j)}$, derived as eigenvectors of the matrix $\Delta_n$ for $n=100$ emitters. Plots in panels (a)-(b) are \textit{acoustic modes}, interpolated by sine functions with frequency $j/(n+1)$, while plots in panels (c)-(d) are \textit{optical modes}, oscillating between a sine function with frequency $(n+1-j)/(n+1)$ and its opposite. The amplitude vectors $\bm{a}^{(j)}$ are normalized to their square norms.
	}
	\label{fig:sinusoid}
\end{figure*}

Figure~\ref{fig:sinusoid} shows the excitation waves $\bm{a}^{(j)}$ for $n=100$ emitters and selected values of $j$: while for small $j$ the excitation profiles can be best understood as the sampling of a sinusoidal function with wavelength $\lambda_j$, with in-phase consecutive amplitudes (\textit{acoustic} waves), for high values of $j$ the components of $\bm{a}^{(j)}$ have alternating signs and values that lie alternatively on two opposite sinusoidal functions with period $2d (n+1)/ (n+1-j)$ (\textit{optical} waves)

\end{document}